%% file: main.tex
\documentclass[%
 twocolumn,
 amsmath,amssymb,
 aps,
prb,
]{revtex4-2}

\usepackage{graphicx}
\usepackage{dcolumn}
\usepackage{bm}

\usepackage{mathtools}
\usepackage{amssymb}
\usepackage{amsmath}
\usepackage{amsthm}
\usepackage{graphicx}
\usepackage{bm}
\usepackage[dvipsnames]{xcolor}
\usepackage{epsfig}
\usepackage{amsfonts}
\usepackage{bbold}
\usepackage{multirow} 
\usepackage{footnote}
\usepackage{blkarray}
\makesavenoteenv{tabular}
\makesavenoteenv{table}
\usepackage[]{appendix}
\usepackage{cancel}
\usepackage[]{hyperref}
\usepackage{soul} 

\newcommand{\expSTMESR}{~\cite{Baumann_Paul_science_2015,
Natterer_Yang_nature_2017,Choi_Paul_natnano_2017,Willke_Paul_sciadv_2018,Yang_Bae_prl_2017,
Willke_Bae_science_2018,Y_Bae_advanced_science_2018,Willke_Singha_nanolett_2019,
Willke_Yang_natphys_2019,Yang_Paul_prl_2019,yang_coherent_2019,Seifert_Kovarik_eabc_2020,
Weerdenburg_Steinbrecher_2020,Steinbrecher_Weerdenburg_2020}}

\begin{document}
\preprint{APS/123-QED}

\title{A study of all-electric electron spin
resonance using Floquet quantum master equations} 

\author{Jose Reina-G{\'{a}}lvez}
\affiliation{Centro de F{\'{i}}sica de Materiales, CFM/MPC  (CSIC-UPV/EHU), 20018 Donostia-San Sebasti\'an, Spain and Center for Quantum Nanoscience, EWHA Womans University, Seoul, Republic of Korea}
\email{galvez.jose@qns.science}
\author{Nicol{\'a}s Lorente}
\affiliation{Centro de F{\'{i}}sica de Materiales
        CFM/MPC (CSIC-UPV/EHU),  20018 Donostia-San Sebasti\'an, Spain}
\affiliation{Donostia International Physics Center (DIPC),  20018 Donostia-San Sebasti\'an, Spain}
\email{nicolas.lorente@ehu.eus}
\author{Fernando Delgado}
\affiliation{Departamento de F{\'{i}}sica, Instituto Universitario de Estudios Avanzados en Física Atómica, Molecular y Fotónica (IUDEA), Universidad de La Laguna 38203,  Tenerife, Spain}
\email{fernando.delgado@ull.edu.es}
\author{Liliana Arrachea}
\affiliation{International Center for Advanced Studies, ECyT-ICIFI, Universidad de San Mart{\'{\i}}n, Av. 25 de Mayo y Francia (1650) Buenos Aires, Argentina}
\email{larrachea@unsam.edu.ar}
\date{\today}

\begin{abstract}
We present a theoretical framework to describe experiments directed to controlling single-atom spin dynamics by electrical means
using a scanning tunneling microscope. We propose a simple model consisting of a quantum impurity connected to electrodes where an electrical time-dependent 
bias is applied. We solve the problem in the limit of weak coupling between the impurity and the electrodes by means of a quantum master equation that is derived by the
 non-equilibrium Green's function formalism. 
 We show results in two cases.
The first case is just a single atomic orbital subjected to a time-dependent electric field, and the second
case  consists of a  single atomic orbital coupled to a second spin-1/2. The first case  reproduces the main experimental features Ti atoms on MgO/Ag (100) while the second one
directly addresses the experiments on two Ti atoms. These calculations permit us to explore the effect of different parameters on the driving of the atomic spins as well as to reproduce experimental fingerprints.
\end{abstract}

\maketitle

\section{Introduction}

Electron transport on the atomic scale has experienced a fast evolution in the last
two decades both experimentally \cite{dermolen_charge_2010} and theoretically \cite{Ferdinand}. The development
of scanning probes together with break junctions has permitted researchers 
to have controlled atomic devices. The explored non-equilibrium electron
transport phenomena open venues for the creation of new technologies \cite{joachim_molecular_2005}.

A prominent example 
is the electron  spin resonance (ESR).
This is a very well-established analysis tool \cite{Abragam_Bleaney_book_1970} routinely
 applied to a variety of fields such as medicine, chemistry and engineering. 
 Standard ESR is an ensemble technique~\cite{Fratila_Velders_arac_2011} with typical commercial devices requiring samples as small as 
100 $\mu$m$^3$.  Recent STM studies~\cite{Mullegger,Baumann_Paul_science_2015} have
pushed ESR to the atomic limit, providing 
a new experimental method capable of spatial atomic resolution and 
nano-eV-energy resolution. This was possible thanks to the integration of ESR and STM, giving place to the STM-based electron spin resonance (STM-ESR). 
In contrast to standard ESR, the new technique is all-electrical where the STM's tip-sample
bias is modulated in time. 

The mechanism for an all-electrical ESR is still under debate \cite{Delgado_Lorente_pss_2021}.
The experiments consist in using a tunneling current produced under a time-dependent applied bias that is focused on magnetic atoms 
with sub-atomic precision. The atomic magnetic moment starts to precess under the effect of the time-dependent electric field of the junction and eventually produces Rabi oscillations
between two magnetic levels distant by a few GHz. When the Rabi oscillations take place, the DC current of the tunneling junctions shows a fast change
with the frequency of the applied bias. This scenario has been unravelled by comparing the DC current with the spin populations obtained from Bloch equations~\cite{Baumann_Paul_science_2015}. This description predicts a Lorentzian-like profile with frequency of the changing DC current, and the strength of the Lorentzian proportional to the square of the Rabi frequency. Other quantities entering the Lorentzian expression otbained from Bloch theory are the lifetimes, $T_1$
of the spin states and the decoherence time, $T_2$. However, the measured peaks actually show asymmetries, that in some cases strongly resemble non-symmetrical Fano
line-shapes~\cite{Yang_Bae_prl_2017,Willke_Paul_sciadv_2018}. More importantly, recent experiments can address several impurities and show collective behavior~\cite{Y_Bae_advanced_science_2018}. A C-NOT two-qubit gate was created by manipulating the system with applied bias pulses~\cite{yang_coherent_2019}.

The recent experimental developments
in STM-ESR\expSTMESR \ call for a suitable  general theoretical framework addressing time-dependent transport in the
presence of local magnetic moments~\cite{Delgado_Lorente_pss_2021}. 
Such a theoretical approach should provide a 
description of the adsorbed molecule, which in general contains multiple orbitals with many-body interactions in a non-equilibrium environment with time-dependent driving. In addition it is also crucial to solve the problem with a reliable many-body technique that correctly 
describes the non-equilibrium conditions.

The above scenario is in the paradigm  of open quantum systems. When a few-level quantum system
is weakly coupled to the environment, it is usual to adopt a description based on quantum master 
equations ~\cite{Breuer_Petruccione_book_2002, Cohen_Grynberg_book_1998,Ribeiro1}. Examples can be found in the study of 
 qubits ~\cite{Nakajima_Noiri_prx_2020,yang2019achieving} or single-dopands in silicon~\cite{vandersypen2017interfacing,Morello_Pla_aqt_2020}
as well as in 
time-dependent electron transport through quantum dots ~\cite{Engel_Hans_prl_2001,Engel_Loss_prb_2002,Ribeiro2,Cavaliere_2009,cava-09,Delgado_Rossier_pss_2017} and other
quantum structures with periodic  driving coupled to reservoirs~\cite{Grifoni_Hanggi_2021}. This consists in formulating the equation of motion of the reduced density matrix of the weakly coupled few-level quantum system either on phenomenological grounds or 
derived from a microscopic Hamiltonian.  
The Green's function formalism (Schwinger-Keldysh)  \cite{rammer_2007} is a well established and reliable method to systematically carry on that derivation in non-equilibrium conditions. 
This formalism is
one of the most widely used in
 quantum transport under time-dependent periodic driving \cite{A_P_Jauho_2006,pastawski1992classical,Jauho_Wingreen_1994,Wingreen,arrachea_green-function_2005} along with Floquet scattering matrix formalism \cite{Moskalets_2002}. In the non-interacting limit there is a one-to-one correspondence between them~\cite{arrachea-moskalets-2006}. 
The derivation of the master equation in the Green's function formalism focuses on calculating
the equation of motion of the non-equilibrium 
reduced density matrix. In our work, the coupling between the system and the reservoirs are treated perturbatively up to second order. 
This was introduced in Ref. \cite{Schoeller,schoen-94} and used in many other works in the context of steady-state \cite{koenig-96-1,koenig-96-2,Esposito_Galperin_prb_2009} and time-dependent transport in the slow  (adiabatic) \cite{spletts-06,Bibek_Bhandari_2021_nonequilibrium} and non-adiabatic \cite{cava-09} regimes. 

\begin{figure}
\includegraphics[width=1.\linewidth]{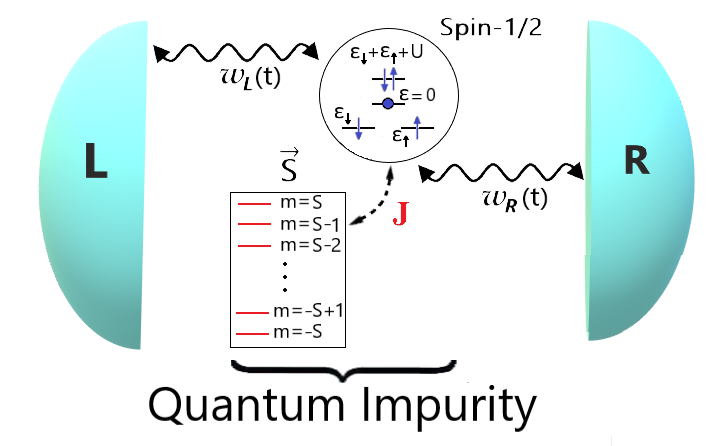}
\caption{Scheme of the model. A single orbital with a Coulomb interaction $U \rightarrow \infty$ is connected via hopping elements, $w_L(t)$ and $w_R(t)$, to the left
and right free-electron electrodes at which an AC voltage of frequency $\omega$ is applied. Emission/absorption of an integer number of a Floquet mode with energy $\ \hbar \omega$ each is associated to this process. In the sketch, the single-orbital is
represented by the circle. When it is occupied by a single electron, it has  spin 1/2 and can be coupled via exchange, $J$, to another spectator spin $S$, represented in the box, that is not tunnel-coupled to the electrodes. The quantum
system in the singly-occupied configuration will then have states with total spin $S_T=S\pm1/2$. The solution based on the Schwinger-Keldysh approach allows us 
to  treat all the quantum degrees of freedom.}
\label{General_scheme}
\end{figure}

The aim of the present work is to study a setup of relevance for ESR-STM experiments \cite{yang_coherent_2019,Veldman,Phark,Mullegger,Baumann_Paul_science_2015}.
 This consists in a quantum impurity (QI) contacted to two electron reservoirs, which represent the substrate and the STM, at which an AC-bias voltage is applied. A sketch is presented in Fig. \ref{General_scheme}. We consider impurities containing internal fully occupied orbitals with a strong Hund coupling, which can be effectively represented by a spin $S$ and an external orbital that hybridizes with the two reservoirs. We focus  on the limit of weak coupling between the impurity and the reservoirs and rely on the non-equilibrium Green's function formalism to derive the quantum master equation. We consider the Floquet representation in the time dependence, which emerges naturally as a consequence of the periodic driving. Our results  suggest that the ingredients considered in our model capture the mechanisms behind ESR-STM experiments.

The article is organized as follows. In Sec.~\ref{Model_Hamiltonian} we introduce the model and the theoretical approach. We derive the equation of motion for the density matrix with the rates expanded in Floquet modes, as well as the equations to calculate the time-dependent current through the device.
Parts of this section are devoted to find explicit  relations between these rates and  the experimentally measured times $T_1$ and $T_2$ as well as the effective Rabi frequency induced by the electrical AC driving.
In Sec. III we present results for the case of an impurity with maximum total spin $S_T=1/2$ while results for larger $S_T$ are presented in Sec. IV. Both of them are computed in the sequential tunneling regime, assuming that  the transport  is dominated by processes where the electrons hop from one of the reservoirs to the system and then to the other reservoir \footnote{By contrast, in STM-ESR, a cotunneling regime applies where the impurity's orbital lies out of the bias window and the current proceeds through the resonance's tails caused by the couplings to the electrodes \cite{Delgado_Rossier_prb_2011,J_Reina_Galvez_2019}. It would require to go to higher order in the hopping.}.
In these sections we relate the predictions based on the Floquet master equation with the features typically analyzed in experiments, like the width of the Lorenztian and Fano-type features observed in the DC-current as a function of the bias voltage and the induced Rabi frequency. The first case  reproduces the main experimental features Ti atoms on MgO/Ag (100) while the second one directly addresses the experiments on two Ti atoms.
Section V is devoted to summary and conclusions. 

\section{Theoretical approach \label{sec_theory}}

\subsection{Model \label{Model_Hamiltonian}}

The  quantum system we study
consists of a quantum magnetic impurity tunnel-coupled to two electron reservoirs, see Fig. \ref{General_scheme}.
We model the impurity's internal structure by a local quantum spin $S$ which represents fully occupied orbitals with a strong Hund rule. This is
exchange-coupled to the spin of an electron trapped in a single external orbital  of the 
impurity. The latter orbital is itself hybridized with the left and right electrodes, which represent the substrate and the tip.
The full system is described by the following Hamiltonian,
\begin{equation}\label{eq:ht}
H(t)=H_{\rm res}+H_{\rm T}(t)+H_{\rm impurity},
\end{equation}
where the first term describes the reservoirs modeled by free electron gases
\begin{equation}\label{eq:hres}
H_{\rm res}=\sum_{\alpha k\sigma }\varepsilon_{\alpha k}c^{\dagger}_{\alpha k \sigma }c_{\alpha k \sigma }.
\end{equation}
Here, $\alpha$ identifies the electrode ($\alpha=$ L,R), while $\sigma =\uparrow, \downarrow$ is the electron spin projection along the quantization axis (chosen along the spin polarization direction of the tip which in general does not coincide with the external magnetic field direction).
These two systems have temperatures and chemical potentials
$T_{\alpha},~\mu_{\alpha}$, respectively. The impurity is modeled by the following Hamiltonian
\begin{eqnarray}\label{eq:himp0}
H_{\rm impurity}&=& \sum_{\sigma} \varepsilon_\sigma d^{\dagger}_{\sigma }d_{\sigma }+ J \mathbf{\hat{s}}\cdot\mathbf{\hat{S}} +D(\hat{S}_z+\hat{s}_z)^2
\nonumber \\
&+&U \hat{n}_{d \uparrow} \hat{n}_{d,\downarrow} + H_{\rm Zeeman},
\end{eqnarray}
where $\varepsilon_\sigma=\varepsilon$ is the energy of the localized  electronic state, 
$U$ is the corresponding Coulomb repulsion, and $\hat{n}_{d\sigma}=d^{\dagger}_{\sigma}d_{\sigma}$ is the occupation operator of the localized state. The localized electrons of the internal orbitals are represented by the spin operator $\mathbf{\hat{S}}$.  
The second term describes the magnetic exchange with strength $J$ between the electron in the hybridized orbital with spin $\mathbf{\hat{s}}$ and the localized spin $\mathbf{\hat S}$.
The spin operator for the electrons on the external orbital has components
$\hat{s}^j=\sum_{\sigma,\sigma'}d^{\dagger}_{\sigma} \hat{\sigma}^j_{ \sigma \sigma'}d_{\sigma'}/2$, being $\hat{\sigma}^j, ~j=x,y,z$  Pauli matrices. Here $D$ is the anisotropy parameter, which is assumed to act on the total spin. For a review on spin Hamiltonians see
for example Ref. [\onlinecite{spinchains}]. The last term of Eq. (\ref{eq:himp0}) represents the Zeeman interaction between the total spin of the impurity, $\mathbf{\hat{S}}_{\rm T}=\mathbf{\hat{S}}+\mathbf{\hat{s}}$and an external uniform magnetic field $\mathbf{B}$, which reads
\begin{equation} \label{h-zeeman}
     H_{\rm Zeeman}=g\mu_B \mathbf{{B}}\cdot(\mathbf{\hat{S}}+\mathbf{\hat{s}}).
\end{equation}
The tunnel coupling between the impurity and the two reservoirs is described by the Hamiltonian 
\begin{equation}\label{eq:ht0}
H_{T}(t)=\sum_{\alpha k\sigma}\left(w_{\alpha }(t)c^{\dagger}_{\alpha k\sigma }d_{\sigma}+w_{\alpha  }^*(t)d^{\dagger}_\sigma c_{\alpha k\sigma }\right).
\end{equation}
In the present treatment, the driving is represented by 
hopping terms, $w_\alpha (t)$, as shown in Refs.~\cite{J_Reina_Galvez_2019,Wolf_C_and_Delgado_F_2020}. 
We consider a tunneling parameter that depends on time as follows
\begin{equation}
w_{\alpha  }(t)=w_{\alpha }^0 \left[1+A_{\alpha }\cos (\omega t) \right], \label{hopping}
\end{equation}
This functional dependence is motivated by the fact that an AC voltage of the form $V_{\alpha} \cos(\omega t+ \varphi_{\alpha})$ applied at
the reservoir $\alpha$ can be expressed after a gauge transformation as a time-dependent tunneling rate $w_{\alpha  }(t)=w_{\alpha}^0 \exp\{-i \int_{t_0}^t dt_1 e V_{\alpha} 
\cos(\omega t + \varphi_{\alpha}) \}$. This function can can be expanded as a Fourier series with Fourier coefficients given by Bessel functions~\cite{Wingreen,arrachea-moskalets-2006}, which for low-enough $eV_{\alpha}/\hbar \omega$ leads to an expression like
Eq. (\ref{hopping}) after a Taylor expansion. Although for arbitrary ratios between these energies it is necessary to consider higher harmonics of the Fourier expansion, for sake of simplicity
we will focus on the simple case of the first-harmonic component as expressed in Eq. (\ref{hopping}). On the other hand, this functional dependence is also motivated by a calculation of the modulation of the tunneling amplitude because of
the applied bias within a WKB approach \cite{J_Reina_Galvez_2019}.
The actual value of the driving amplitude $A_{\alpha}$ depends on the modulation of the hopping. Particularly, one obtains $A_\alpha$ ranging from $10^{-3}$ to $0.5$ for the typical STM-ESR experimental conditions~\cite{J_Reina_Galvez_2019}.

We focus on the limit of $U \rightarrow \infty$ in Eq. (\ref{eq:himp0}), which corresponds to  the situation where the external orbital can be only empty or singly occupied. In such a case, it is convenient to introduce the basis
$|p,m\rangle $,  where the first entry $p=0,\uparrow, \downarrow$ corresponds to the state of the conduction orbital while the second one corresponds the magnetic quantum number $m=-S, \ldots, S$ associated to the localized spin. The Hamiltonian for the impurity defined in Eqs. (\ref{eq:himp0}),  in the limit of vanishing double occupation, can be written as  
\begin{eqnarray}\label{eq:himp}
H_{\rm S}&=&\sum_{p, m}\varepsilon_{pm} |p,m\rangle \langle p,m|+H_{\rm Zeeman} \nonumber
\\
&&+
\sum_{m}^{S-1} \left(J_m |\downarrow,m+1\rangle \langle \uparrow,m|+ h.c. \right), 
\end{eqnarray}
where we have introduced the definitions
\begin{eqnarray}
\varepsilon_{pm} & =&\delta_{p,\sigma }\left[\varepsilon + Jm\frac{s_{\sigma}}{2}  +D\left(\frac{1}{4} + m s_{\sigma}\right) \right] + D m^2,
\crcr
J_m&=&\frac{J}{2} \sqrt{S(S+1)-m(m+1)},
\end{eqnarray}
with $s_{\uparrow}= +1\; (s_{\downarrow}= -1)$. 
The corresponding tunneling Hamiltonian expressed in this basis reads 
\begin{equation}\label{ht}
H_T'(t)=\sum_{\alpha k \sigma m}\left(w_{\alpha }(t)c^{\dagger}_{\alpha k\sigma } |0,m\rangle \langle \sigma,m|  + h.c. \right).
\end{equation}
It is convenient to express the Hamiltonian for the impurity and the tunneling coupling in the basis of eigenstates of the Hamiltonian defined in Eq. (\ref{eq:himp}), $H_{\rm S} |l\rangle = E_{l} |l \rangle, ~l = 1, \ldots, 3(2 S+1)$. Accordingly, $H_T'(t)$ is written as follows
\begin{equation}
    H_T'(t)=\sum_{\alpha k, \sigma, lj} \left(w_{\alpha }(t)c^{\dagger}_{\alpha k\sigma } \lambda_{l j,\sigma} |l\rangle   \langle   j|  + h.c. \right),
\end{equation}
with 
\begin{equation}
    \lambda_{lj,\sigma} = \sum_{ m} \langle l|  0,m\rangle \langle \sigma,m |j \rangle.
\end{equation}
Notice that the Hamiltonian of Eq. (\ref{eq:himp}) commutes with the particle-number operator for the external orbital, $\hat{n}_d=\sum_{\sigma} d^{\dagger}_{\sigma}d_{\sigma}$.
Hence, the states $|l\rangle$ are states with a well defined
occupancy number $n_d=0,1$.
The total spin of the isolated impurity for $\mathbf{{B}}=D=0$, is a good quantum number, which takes values
 $S_T=S\pm1/2$, for $S>0$ and $S_T=1/2$ for $S=0$, in the configuration with $n_d=1$.

\subsection{Floquet master equation} \label{dynamics}
We proceed similarly to Refs. [\onlinecite{schoen-94,koenig-96-1,koenig-96-2,spletts-06,Esposito_Galperin_prb_2009,cava-09,Bibek_Bhandari_2021_nonequilibrium}]  to derive the quantum master equation by treating the coupling between the impurity and the reservoirs at the lowest  (second) order in perturbation theory in $H_T$, (equivalent to the Born-Markov approximation \cite{rammer_2007,Dorn_2021}) in the framework of non-equilibrium Green's function formalism.
This is appropriate for the description of the sequential tunneling regime, which corresponds to 
the effective tunneling between the two reservoirs dominated by processes where the electrons hop from one of the reservoirs to the system and then to the other reservoir.  This implies that the hybridization $w_\alpha(t)$ should be small enough to avoid drastic disturbances of the spectra of $H_S$ and that fourth-order processes where the charge of the impurity changes in virtual processes must be negligible. This last condition is satisfied if the chemical potential of the impurity $\mu_{\rm imp}^\alpha(n_d)=E_{\rm GS}(n_d+1)-E_{\rm GS}(n_d)$, being $E_{GS}(n_d)$ the ground state energy of the impurity with $n_d$ electrons,  is close to the chemical potential of the electrodes, i.e., $|\mu_{\rm imp}^\alpha(N)-\mu_{\alpha}|\lesssim k_BT, |eV|$.
Details of the derivation are presented in Appendix \ref{derivation_floquet_master_equation}. 

The derived equation ruling the dynamics of $\rho_{lj}(t)$, which are the elements of the reduced density matrix defined as,
\begin{equation}\label{rho}
\rho_{lj} (t) = \mbox{Tr}\left[\hat{\rho}_{T} (t)  \hat{\rho}_{lj} \right],
\end{equation}
with $\hat\rho_{lj}=|l\rangle\langle j|$ and the trace taken over the degrees of freedom of the total system. $|l\rangle, \; |j\rangle$ denote many-body eigenstates  of the Hamiltonian $H_{\rm S}$. The equation of motion for $\rho_{lj}(t)$ becomes --see details in Appendix \ref{derivation_floquet_master_equation}-- 
\begin{eqnarray}
& &\hbar\dot{\rho}_{lj}(t)- i\Delta_{lj}\rho_{lj}(t)= \sum_{vu}\left[\Gamma_{vl,ju}(t)\rho_{vu}(t)+\bar{\Gamma}_{vl,uv}(t)\rho_{uj}(t) \right.\nonumber \\
& &\left.  ~~~~~~~~~~~- \Gamma_{jv,vu}(t)\rho_{lu}(t)-\bar{\Gamma}_{jv,ul}(t)\rho_{uv}(t)\right],
\label{rho_master_eq_6_0}
\end{eqnarray}
where we have denoted $\Delta_{lj}=E_l-E_j$. The time-dependent driving entails an implicit time-dependence of the rates $\Gamma(t)$ and $\bar{\Gamma}(t)$, which fulfill the relation $\Gamma_{vl,ju}(t)=-\bar{\Gamma}_{lv,uj}^{\ *}(t)$. As a consequence of the AC driving, these periodic rates can be expandend in terms of Fourier-Floquet components as follows,
\begin{eqnarray}
\Gamma_{vl,ju}(t) &=&\sum_{n'} e^{-in'\omega t}\Gamma_{vl,ju;n'}(\omega),
\label{Rate_Floquet_form}
\end{eqnarray}
where $\Gamma_{vl,ju;n'}(\omega)=-\bar{\Gamma}_{lv,uj;-n'}^{\ *}(\omega)$. The explicit calculation of the Fourier-Floquet components of the rates for the model of Sec~\ref{Model_Hamiltonian} leads to
\begin{eqnarray}
\Gamma_{vl,ju;n'}(\omega)&=& \sum_\alpha \left[\Gamma_{vl,ju,\alpha;n'}^0(\omega)- \bar{\Gamma}_{vl,ju,\alpha;n'}^1(\omega)\right],
\label{Rate_Floquet_divided}
\end{eqnarray}
with
{\small \begin{widetext}
\begin{eqnarray}
\Gamma_{vl,ju,\alpha;n'}^0(\omega)&=&\sum_{n\sigma}\frac{ \lambda_{vl\sigma}\lambda^{*}_{uj\sigma} }{2}
\gamma_{\alpha\sigma} f_\alpha(\Delta_{ju}-n\hbar\omega) \left[\delta_{n,n'} +\frac{A_{\alpha}}{2}\left(\delta_{n,n'+1}+\delta_{n,n'-1}\right) \right]\left[\delta_{n,0} +\frac{A_{\alpha}}{2}\left(\delta_{n,1}+\delta_{n,-1}\right)\right] \nonumber
\\
\Gamma_{vl,ju,\alpha;n'}^1(\omega)&=&  \sum_{n\sigma} \frac{ \lambda^*_{lv\sigma} \lambda_{ju\sigma} }{2} \gamma_{\alpha\sigma} (f_\alpha(\Delta_{uj}-n\hbar \omega)-1)  
\left[\delta_{n,-n'} +\frac{A_{\alpha}}{2}\left(\delta_{n,1-n'}+\delta_{n,-n'-1}\right) \right]\left[\delta_{n,0} +\frac{A_{\alpha}}{2}\left(\delta_{n,1}+\delta_{n,-1}\right) \right].
\label{Rate_Floquet_final_appendix}
\end{eqnarray}
\end{widetext}}
We notice that the $n'$ index in the rates components can only take the values -2,-1,0,1,2 as a consequence of the fact that the driving depends on a single harmonic dependence $\cos(\omega t)$ in Eq. (\ref{hopping}) which enters to the square in the rates. Importantly, all the information on the temperature $T_\alpha$ and chemical potential $\mu_{\alpha}$ of a given reservoir is encoded in the Fermi distribution functions, $
f_\alpha(\epsilon)=1/\left(e^{(\epsilon-\mu_{\alpha})/k_B T_\alpha}+1\right)$.  The chemical potential only contains the corresponding DC component of the bias.
For convenience, we define the hybridization function of an electron with spin $\sigma$,
\begin{equation}\label{pol}
    \gamma_{\alpha\sigma}=2\pi\rho_{\alpha\sigma}|w_\alpha|^2=
    \frac{1}{2}\left(1+2\sigma P_{\alpha} \right)\gamma_{\alpha}' ,
\end{equation}
which depends on the spin-dependent density of states $\rho_{\alpha\sigma}$. Here we have also defined the polarization $P_\alpha$ and $\gamma'_{\alpha}=2\pi\rho_\alpha |w_\alpha|^2$. 

We notice that as a consequence of the Floquet structure, the stationary solution of the master equation  depends periodically on time. 
Hence, it is appropriate to expand it as

\begin{equation}
\rho_{lj}(t)=\sum_n e^{-i n \omega t}    \rho_{lj;n},
\end{equation}
with

\begin{equation*}
\rho_{lj;n} = \frac{\omega}{2\pi} \int_0^{2 \pi /\omega} \rho_{lj}(t) e^{i n \omega t}.
\end{equation*}
Substituting in Eq. (\ref{rho_master_eq_6_0}), we get the Floquet master equation
\begin{widetext}
\begin{equation}
 \Delta_{lj}\rho_{lj;n}  + n\hbar \omega  \rho_{lj;n}=i\sum_{vu; n'}\left[\Gamma_{vl,ju;n'}(\omega)\rho_{vu;n-n'}+\bar{\Gamma}_{vl,uv;n'}(\omega)\rho_{uj;n-n'}- \Gamma_{jv,vu;n'}(\omega)\rho_{lu;n-n'}-\bar{\Gamma}_{jv,ul;n'}(\omega)\rho_{uv;n-n'}  \right].
\label{rho_master_eq_Floquet}
\end{equation}
\end{widetext}
The structure of Eq. (\ref{rho_master_eq_Floquet}) reflects the fact that the dynamics of the density matrix depends on the exchange of an integer number of Floquet modes with energy $\hbar \omega$ each, introduced by the hybridization with the driven reservoirs. 

Finally, we also  add the normalization condition. 
Since $\sum_l \rho_{ll}=1 \implies \sum_{ln}e^{-in\omega t} \rho_{ll;n}=1$. If we  multiply by $e^{in''\omega t}$ and integrate over $t$, we obtain $\sum_l \rho_{ll;n}=\delta_{n,0}$.

\subsection{Current through the impurity}

One interesting observable is the current that goes through the impurity as a response to the bias voltage. We can proceed along similar lines as in the derivation of the master equation for the
density matrix. The current flowing out of the lead $\alpha$ is defined as $I_{\alpha}=-e\frac{d\langle N_{\alpha}\rangle}{dt} $, and explicitly reads 
\begin{equation}
I_{\alpha}(t)=
\frac{e}{\hbar}\sum_{k\sigma l j}\left(w_{\alpha}\lambda_{lj\sigma}G_{lj,\alpha k\sigma}^{<}(t,t) - w_{\alpha }^* \lambda^*_{jl\sigma}G_{\alpha k\sigma,lj}^{<}(t,t)  \right).
\end{equation}
Following the same procedure used in Appendix \ref{derivation_floquet_master_equation}, we write
\begin{equation}
I_{\alpha}(t,\omega)=
\frac{2e}{\hbar}\sum_{ lju} \sum_{nn'} e^{-in\omega t}  \mbox{Re} \left[\rho_{lu;n-n'} \Gamma_{lj,ju,\alpha;n'}(\omega) \right].
\label{Current_floquet_simp}
\end{equation}
In the derivation of the previous expression we have defined 
\begin{equation}
\Gamma_{vl,ju,\alpha;n'}(\omega)=\Gamma_{vl,ju,\alpha;n'}^0(\omega)+\Gamma_{vl,ju,\alpha;n'}^1(\omega).
\label{Rates_current}
\end{equation}
By computing the rates and the Floquet density matrix elements we can calculate the current. Experimentally, only the DC current, $n=0$, is accessible and, for the specific time dependent hopping $w_{\alpha}(t) $ defined in Eq. (\ref{hopping}), $n'$ can only take the values $-2,-1,0,1,2$. From here on, we are going to consider the driving to be small so that we can neglect   contributions $\propto A_{\alpha}^2$ in the rates.

\subsection{DC regime: Lifetime and decoherence time}
\label{Relaxation and decoherence time}
We now focus on the steady state limit, which corresponds to $A_{\alpha}=0$. Therefore, the stationary rates intoduced in Eq. (\ref{Rate_Floquet_form})
reduce to $\Gamma_{vl,ju}(t) \equiv \Gamma_{vl,ju;0}$ and  $\bar{\Gamma}_{vl,ju}(t)\equiv \bar{\Gamma}_{vl,ju;0}$. In this case, we can proceed
as in the treatment of the Bloch-Redfield equation to define the lifetime of a given state as well as the dephasing time \cite{Delgado_Rossier_pss_2017}. We summarize below the main steps.

In order to define  the lifetime we focus on Eq. (\ref{rho_master_eq_6_0}) for the diagonal elements of the density matrix -- {\em populations} -- assuming that the non-diagonal ones -- {\em coherences} -- do not contribute. By using the property  $\Gamma_{lj,vu;0}=-\bar{\Gamma}_{jl,uv;0}$, the corresponding equation reads
\begin{eqnarray}
& &\hbar\dot{\rho}_{ll}(t)= 2 \sum_{v\neq l}\left[\Gamma_{vl,lv;0}\rho_{vv}(t)-\Gamma_{lv,vl;0}\rho_{ll}(t) \right],
\label{rate_master_eq_6_1}
\end{eqnarray}
where we identify the inverse of the population life time of the state $|l\rangle$ by 
\begin{equation}\label{life}
\frac{1}{T_{1}^l}= \frac{2}{\hbar} \sum_{v\neq l} \Gamma_{lv,vl;0}.
\end{equation}
Notice that if the state $|l\rangle$ corresponds to a many-body state with single-electron occupancy, $v$ runs over all the states with zero-electron occupancy. 

Similarly, if we now focus on the coherences of Eq. (\ref{rho_master_eq_6_0}) we get
 \begin{eqnarray}
\hbar\dot{\rho}_{lj}(t)- i\Delta_{lj}\rho_{lj}(t)&=&-\sum_{v}\left(\Gamma_{jv,vj;0}+\Gamma_{lv,vl;0}\right)\rho_{lj}(t)+\ldots \nonumber
\\
&&\label{coh_master_eq_6_0}
\end{eqnarray}
where $l\neq j$ and correspond to singly-occupied states while $\ldots$ denotes the other terms of the master equation: coherences $\rho_{uv}$ with $u\neq l$ and $v\neq j$ plus population terms. The sum over  $v$ runs over empty states. 
As 
the rates entering this equation correspond to second-order tunneling processes between the impurity and the reservoirs, coherences with states $l$ corresponding to an empty state and $j$ corresponding to a singly-occupied one or vice-versa are zero.
Eq. (\ref{coh_master_eq_6_0}) leads  to the following definition of the inverse of the coherence time
\begin{equation}
\label{coherence}
    \frac{1}{T_2^{lj}}=\frac{1}{\hbar}\sum_{v}(\Gamma_{lv,vl;0}+\Gamma_{jv,vj;0}).
\end{equation}
Hence we can identify a relation between the coherence and the relaxation times
\begin{equation}
    \frac{1}{T_2^{lj}}=\frac{1}{2}\left(\frac{1}{T_1^l}+\frac{1}{T_1^j}\right).
\end{equation}
 We notice the lack of pure dephasing time in this description. This is because, for the type of bath we are considering, the fluctuation between two singly occupied states of the impurity necessarily implies a second-order charge fluctuation. Hence, the rates of the form $\Gamma_{ll,jj}$ are zero.

\subsection{AC regime: Rabi frequency \label{Rabi frequency}}

The Rabi frequency of a magnetic system under a static magnetic field and driven by time-dependent transverse magnetic field
is  defined 
as the transfer rate of the population of one state towards the other state induced by the periodic driving.
The latter is represented
by a time-oscillating term in the 
 off-diagonal matrix elements of the Hamiltonian 
 expressed in the basis of the spinors in the direction of the static magnetic field. 
 
Contrary to the standard ESR~\cite{Abragam_Bleaney_book_1970}, here we are not considering a time-dependent periodically driven magnetic field. In fact, the only time-dependent term of our model is in the form of the periodically modulated hybridization of the impurity level and the leads, see Eq. (\ref{hopping}). 
 We analyze now the possibility of inducing an oscillating transfer rate of the populations of the different states of the impurity, akin to the Rabi oscillations, as a consequence of the electrical AC bias.
 
 We focus on $|l\rangle$ being a singly-occupied state of the impurity. The evolution of the corresponding population is given by
 \begin{eqnarray}\label{rholt}
\hbar \dot{\rho}_{ll}(t)&=& 2\sum_{v}\left\{\mbox{Re}[\Gamma_{vl,lv}(t)]\rho_{vv}(t)- \mbox{Re}[\Gamma_{lv,vl}(t)]\rho_{ll}(t) \right\} +
\nonumber \\
&&\sum_{vu;v\neq u}\left[\Gamma_{vl,lu}(t)\rho_{vu}(t) - \bar{\Gamma}_{lv,ul}(t)\rho_{uv}(t)\right]  \nonumber -
\\
&&\sum_{vu;l\neq u}\left[ \Gamma_{lv,vu}(t)\rho_{lu}(t) -\bar{\Gamma}_{vl,uv}(t)\rho_{ul}(t) \right],
\end{eqnarray}
where, as already mentioned $v$ runs over empty states  for $l$ labeling a singly-occupied one.

We extend the previous definition of Eq. (\ref{life}) of the inverse of the life time to the present case, where we are considering time-dependent rates, as follows
\begin{equation}
\frac{1}{T_{1}^l(t)}=\frac{2}{\hbar}\sum_{v;v\neq l}\mbox{Re}[\Gamma_{lv,vl}(t)],~~~~~~\frac{1}{T_{1}^{vl}(t)}=\frac{2}{\hbar}\mbox{Re}[\Gamma_{vl,lv}(t)].
\end{equation}
We have also introduced the inverse of the life time $T_{1}^{vl}(t)$.
We identify the terms connecting coherences as  Rabi-type terms,
\begin{equation}
\frac{\hbar\Omega_{vu}^l(t)}{2} = \Gamma_{vl,lu}(t)=-\bar{\Gamma}_{lv,ul}^{\ *}(t).
\end{equation}
Introducing this definition, Eq. (\ref{rholt}) reads
\begin{eqnarray}
\dot{\rho}_{ll}(t)&= &\sum_v \frac{\rho_{vv}(t)}{T_{1}^{vl}(t)} -\frac{\rho_{ll}(t)}{T_{1}^l(t)}  + \sum_{vu;v\neq u}\mbox{Re}\left[\Omega_{vu}^l(t) \rho_{vu} (t)\right]-
\nonumber\\
&&\sum_{vu;l\neq u}\mbox{Re}\left[\Omega_{lu}^v(t)\rho_{lu}(t)\right].
\end{eqnarray}
The transition amplitudes can be Floquet-Fourier transformed. The Floquet sum can only take values from $-2$ to $2$, see Eq. (\ref{Rate_Floquet_final_appendix}). For
$\omega$ much smaller than the inverse of any other characteristic time of the problem, we can perform the approximation that the Fermi distribution in the rates functions, Eq. (\ref{Rate_Floquet_final_appendix}), do not depend on $\omega$ strongly. 
Under this assumption, plus neglecting terms proportional to $A_\alpha^2$ and considering no coherence at zero driving, Eq. (\ref{rate_equation_rabi}) can be further simplified and reads
\begin{eqnarray}
\dot{\rho}_{ll}(t)&= &\sum_v \frac{\rho_{vv}(t)}{T_{1}^{vl}(t)} -\frac{\rho_{ll}(t)}{T_{1}^l(t)}  +\nonumber
\\
&& \sum_{vu;v\neq u}\Omega_{vu;1}^l [\rho_{vu}(t)+\rho_{uv}(t)]\cos\omega t-
\nonumber\\
&&\sum_{vu;l\neq u}\Omega_{lu;1}^v(t)[\rho_{lu}(t)+\rho_{ul}(t)]\cos\omega t,
\label{rate_equation_rabi}
\end{eqnarray}
which resembles the Bloch-Redfield master equation with the Rabi term. Therefore, we identify from this equation the Rabi frequency of our model 
\begin{eqnarray}\label{rabi}
\hbar \Omega_{lu;1}&=&\hbar \sum_v \Omega_{lu;1}^v= \hbar \sum_v\Omega_{lu,L;1}^v +\hbar \sum_v\Omega_{lu,R;1}^v= 
\nonumber
\\ && 2\sum_{v\alpha} A_{\alpha} \left(\Gamma_{lv,vu,\alpha;0}^0- \Gamma_{lv,vu,\alpha;0}^1 \right).
\end{eqnarray}
This quantity is found to be zero for non-polarized reservoirs ($P_\alpha=0$).

Eq. (\ref{rate_equation_rabi}) can be cumbersome to solve analytically. However, numerical calculations suggest that the populations contribute with a term proportional to the square of  Rabi frequencies, while the coherences have a linear dependence on this quantity. This is also what other studies found employing the rotating wave approximation \cite{Delgado_Rossier_pss_2017,Cohen_Grynberg_book_1998}. Therefore, the DC current, $n=0$ in Eq. (\ref{Current_floquet_simp}),  behaves as $\Delta I_{\alpha}^{DC}(\omega) \propto \sum_{lu,l\neq u}\left[\Omega_{lu;1}\right]^2$ at the corresponding resonance frequencies between filled states $|l\rangle$ and $|u\rangle$.


\input{Resultados}

\section{Summary and conclusions}

We have presented a theory based on non-equilibrium Green's functions introducing the Floquet representation to derive the quantum master equation for the reduced density matrix and the DC current through a quantum impurity weakly coupled to two electrodes with AC electrical driving. This setup is relevant for experiments of STM-ESR where the spin dynamics is induced 
by electrical AC driving.

We considered a  model containing the main ingredients that are expected to play a role in experiments of ESR-STM. These are the exchange interaction between the spin occupying the external orbital of the molecule and the localized effective spin corresponding to the occupied internal orbitals, the spin anisotropy, the polarization of the electrodes and an external magnetic field.  

We studied  two simple cases. The first one corresponds to an impurity with total spin $S_T=1/2$.
We have clearly shown that, under a  driving external electric field, we do have induced Rabi oscillations and that an ESR is established. Our theory has permitted us to define the strength of the Rabi frequency, and correlate it with the height of the resonance. Our results indicate that it is crucial to have polarization in the electrode in order to have Rabi oscillations in this configuration. Moreover, we have computed the coherence time, $T_2$ and shown that is related to the width of the ESR. 
Thus, we have successfully replicated the main experimental features for the single Ti atom, showing the information contained in the line profiles of the ESR. 

In the second case, following the experimental studies of two Ti atoms with STM-ESR \cite{Y_Bae_advanced_science_2018}, we have briefly analyzed a setup which corresponds to a maximum total spin $S_T=1$: coupling of two spins one half. This configuration is richer and leads to several signals in the profile of the DC current as a function of the driving frequency, which are associated to the transitions between different triplet states. In this case, the polarization also plays a fundamental role to generate Rabi frequencies and it can lead to vanish resonance peaks in the half metal case.

The present study can be extended in several directions, including the description of other adsorbates, like Fe, to reproduce the main experimental STM-ESR results for Fe adsorbates ~\cite{Baumann_Paul_science_2015}, where indeed fourth-order Stevens operators are necessary. Another direction is to include the charge fluctuations
related to the doubly occupancy of the external orbital along with a finite value of the Coulomb interaction $U$ (here we have considered $U\rightarrow \infty$). Other extension of the present theory is to consider higher order processes in the tunneling term, which has been considered at the lowest (2nd order in the current) here. This would allow us to explore effects which have not been considered in the present work, such as cotunneling and dephasing processes. This approach can be also extended to analyze noise \cite{caso_model_2014} and is also appropriate for the study of similar systems like  time-dependent spin-orbit couplings in quantum dots \cite{Entin-Wohlman_2020_spin-orbit,Entin-Wohlman_2020_Rashba}.

\begin{acknowledgments}

We are pleased to thank our collaborators for important discussions. A non-exhaustive list of the
many contributors to our discussions is:
D.-J. Choi,
T. Choi,
F. Donatti,
A. J. Heinrich,
J. Fernandez-Rossier,
P. Gambardella,
L. Limot,
F. Natterer,
T. Seifert,
S. Stepanow,
P. Willke,
C. Wolf.
Financial support from the Spanish MICINN (projects RTI2018-097895-B-C44 and PlD2019-109539G8-C41) is gratefully acknowledged. FD acknowledges financial support from Basque Government, grant IT986-16
and Canary Islands program {\em Viera y Clavijo}
(Ref. 2017/0000231). LA is supported by CONICET, and also acknowledges financial support from PIP-2015 and ANPCyT, Argentina, through PICT-2017-2726, PICT-2018-04536, as well as the Alexander von Humboldt Foundation, Germany. 

\end{acknowledgments}

\begin{appendices}
\addappheadtotoc

\input{Appendices}

\end{appendices}

\bibliographystyle{apsrev}
\bibliography{references.bib}

\end{document}

%% file: Resultados.tex
\section{Results for $S_T= 1/2$}
\begin{figure}
\includegraphics[width=1.\linewidth]{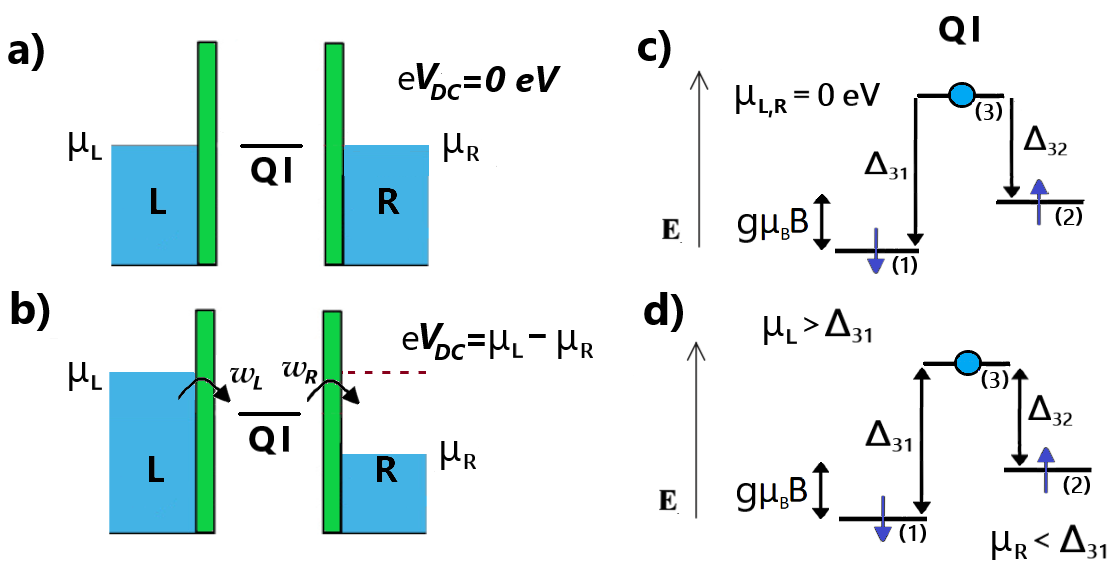}
\caption{Static energy scheme of the $S_T=1/2$ QI Hamiltonian (right) and the cotunneling configurations (left). Plots a) and b) are one-electron energy schemes representing a 
QI system at zero and finite bias, while panels c) and d) show the many-body energies of the corresponding states involved in the tunneling process. The chemical potential for both electrodes is taken as the reference level, $\mu_{L,R}=0$. At zero-bias, a) and c), no net current is passing through the QI. 
 In the QI configurations at zero magnetic field $|1\rangle\equiv|\downarrow\rangle$ and $|2\rangle\equiv |\uparrow\rangle$.
State $|3\rangle$ is the empty state of the QI.
Here $\varepsilon_{\sigma}<0$ such that the up and down states are energetically below the empty one. 
In order to induce a current through the QI and arrive to pictures b) and d), we apply a DC bias voltage such that $eV_{DC}=\mu_L-\mu_R>0$, by  decreasing $\mu_R$ while, symmetrically, increasing $\mu_L$ until the absolute value of the right Fermi energy level is larger than the difference $\Delta_{31}=-\varepsilon_{\downarrow}-g\mu_B |\mathbf{B}|/2$. Similarly, $\Delta_{32}=-E_2=-\varepsilon_{\uparrow}+g\mu_B |\mathbf{B}|/2$.  }
\label{Scheme_S=0}
\end{figure}

The simplest case corresponds to  an impurity with $S=0$ for the localized spin degrees of freedom. Hence, the total spin when the external orbital is singly occupied is $S_T=1/2$.
This configuration is equivalent to a quantum dot with large charging energy ($U\rightarrow +\infty$), which has only three many-body states, corresponding to two occupied levels (spin up and down) and an empty level, as shown in the scheme of Fig. \ref{Scheme_S=0}.

\subsection{Model parameters \label{Parameters}}

To illustrate the workings of our ESR mechanism, we fix the model parameters so that the system is in the sequential tunneling regime. This makes the analysis particularly simple.  In addition, we can expect the main features of the resonant ESR signal to remain when the current is dominated by off-resonance transport (cotunneling regime)~\cite{Engel_Loss_prb_2002}.
The parameters used for the calculations of this
section are as follows unless otherwise specified: 
\begin{itemize}
\item The magnetic field is fixed to $\mathbf{B}=(5.0,0,0.2)T$, where the tip's spin polarization is in the $z$ axis. 

\item We particularize to a weak driving regime, as expected in the all-electrical STM-ESR implementations~\cite{J_Reina_Galvez_2019}. Thus, we take $A_\alpha\in [10^{-3},0.5]$.

\item The driving constants are  $A_L=A_R=0.05$ and the hybridization
functions, $\gamma'_{\alpha}=2\pi\rho_\alpha |w_\alpha|^2$, $\gamma'_L=\gamma'_R=50$ $\mu$eV. The right electrode is spin polarized with {$P_R=0.35$.}

\item The electronic level is set to $\epsilon_\downarrow=\epsilon_\uparrow=-1$ meV, fixing the energy difference between electronic states to $\Delta_{31}= 1.29 $ meV and $\Delta_{32}=0.71$ meV, see Fig.~\ref{Scheme_S=0}.

\item The system of equations in (\ref{rho_master_eq_Floquet}) are limited to $n=5$. This Floquet number controls the convergence of our results and a value of 5 suffices our purpose.

\item The temperature is fixed at one Kelvin.

\end{itemize}

\subsection{Static case, $\omega=0$}
\begin{figure}
\centering
\includegraphics[width=0.8\linewidth]{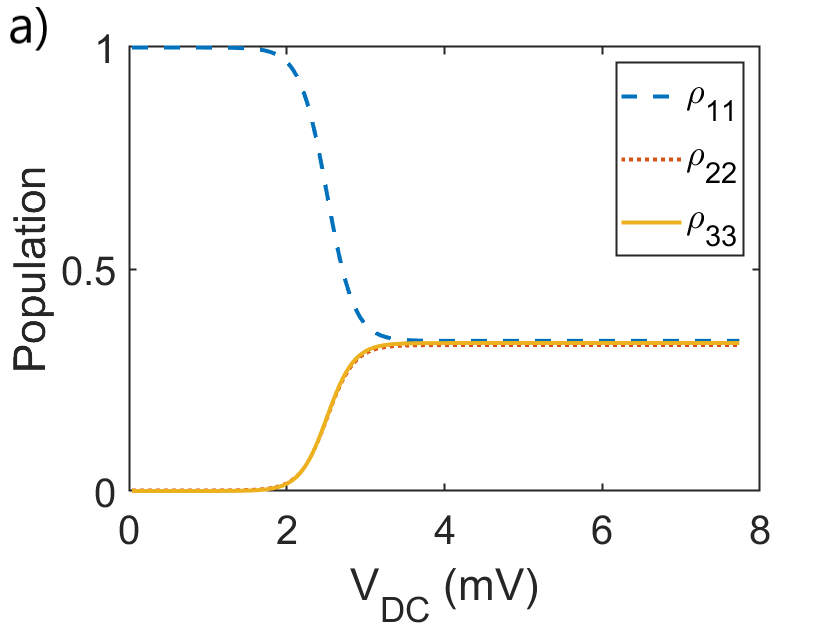}
\includegraphics[width=0.87\linewidth]{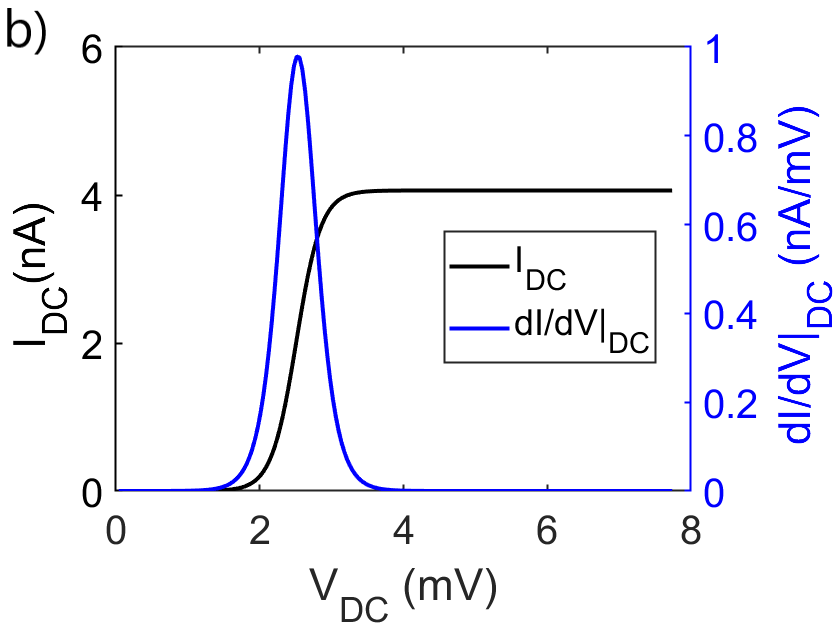}
\caption{a) Population vs DC voltage and b) current and conductance as function of the bias. The saturation in the current indicates that we are in the sequential tunneling regime where the populations are 1/3 at zero driving and spin polarization. The parameter used are indicated in~\ref{Parameters}. }
\label{rhos_I_vs_V}
\end{figure}

We  analyze now the simplest situation where the AC bias voltage is not applied, $A_\alpha=0$.
We start by considering both electrodes at zero chemical potential $\mu_{\alpha}=0$, as indicated in 
Fig. \ref{Scheme_S=0}a) and c). Next, we change symmetrically the chemical potentials of the electrodes with respect to this value, $\mu_L=-\mu_R=e V_{\rm DC}/2$.  
The resulting populations are plotted as a function of the applied bias in Fig. \ref{rhos_I_vs_V}a). At low bias, the ground state $|1\rangle$ is completely populated.
When the bias potential matches $2\Delta_{31}$, the chemical potential of the dot becomes resonant with the chemical potential of the right electrode, allowing tunneling of the down electron to the right electrode. Moreover, since $\Delta_{32}\le \Delta_{31}$, the transport channel through state $|2\rangle$ will also be opened.
As a consequence, the three states of the QI become equally populated ($\rho_{jj}\approx 1/3)$).
The response to the voltage bias is also revealed in the behavior of the current, see Fig. \ref{rhos_I_vs_V}b). 
As the bias overcomes the value $eV_{DC}=2\Delta_{31}$, the current increases  reflecting the opening of a transport channel, while 
it saturates at a higher bias when all channels are fully open. Correspondingly, the $dI/dV$ peaks at the current step bias $eV_{DC}=2\Delta_{31}$.

\subsection{AC-driven case, $\omega\neq 0$}
\label{AC_driven}

\begin{figure}
\centering
\includegraphics[width=0.8\linewidth]{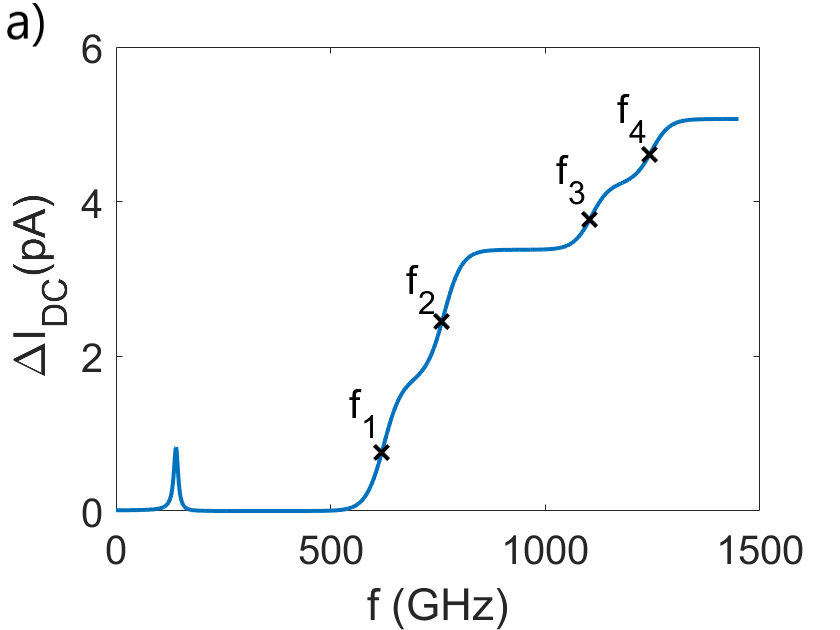}
\includegraphics[width=0.8\linewidth]{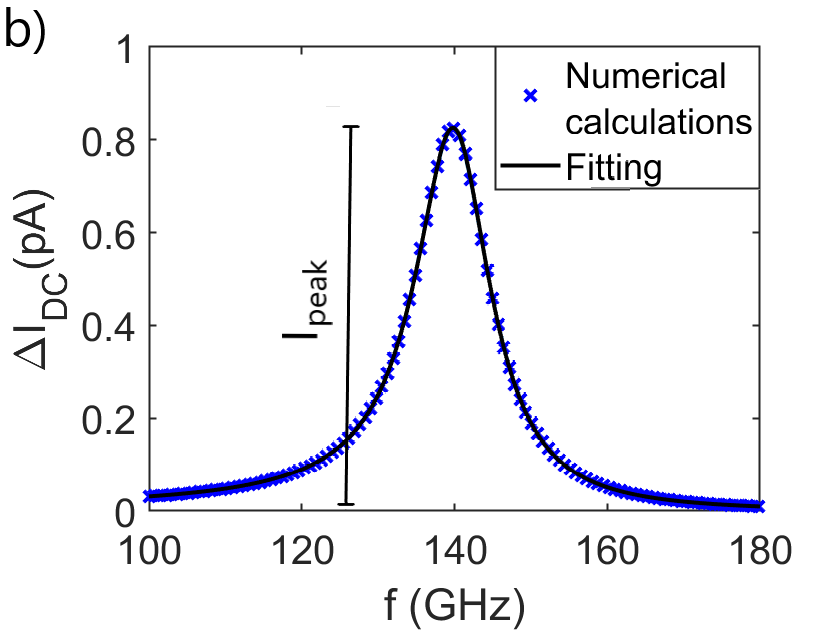}
\caption{Change in the current $\Delta I_{DC}(f)$ vs frequency for a fixed DC bias voltage of $7.7$ mV ($\mu_L=-\mu_R=3.85$ meV) and the set of parameters introduced in Sec.~\ref{Parameters}.   
For the displayed frequency interval, $10 \hbar\omega < e V_{DC}$, 
so the frequency does not alter the Fermi occupations under adiabatic conditions. The whole spectrum is depicted in panel a)  while a zoom of the resonance peak is plotted on panel b). A fit to a Fano line-shape is shown in panel b) too.
The resonance peak position coincides with the energy difference between states $|1\rangle$ and $|2\rangle$ which is $g \mu_B |\mathbf{B}|\approx 5.79\times 10^{-1}$ meV or $140$ GHz. 
The other transitions, $f_i$, are identified in the text. When comparing with the background longtime averaged current,
the relative change in current at resonance is -0.02\%.
}
\label{I0_vs_omega}
\end{figure}

\begin{figure}
\centering
\includegraphics[width=0.8\linewidth]{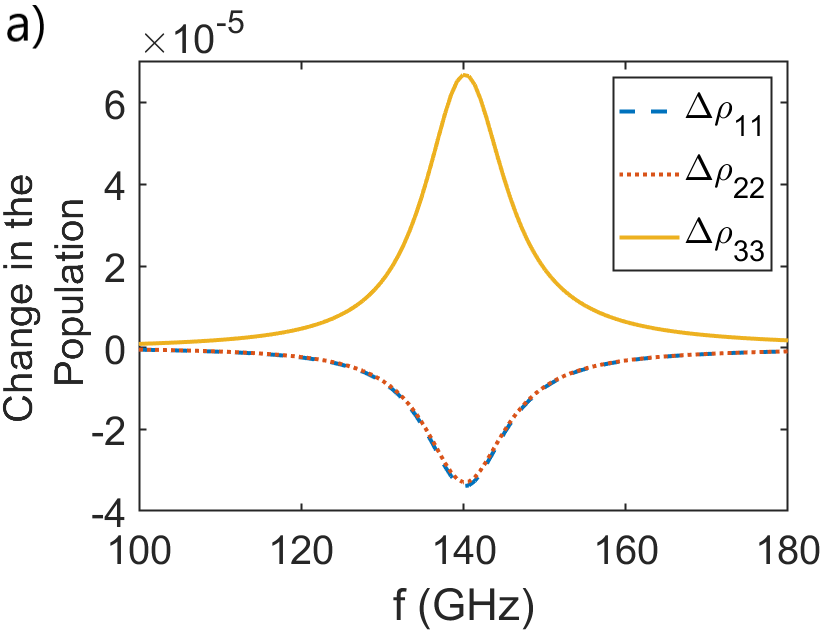}
\includegraphics[width=0.8\linewidth]{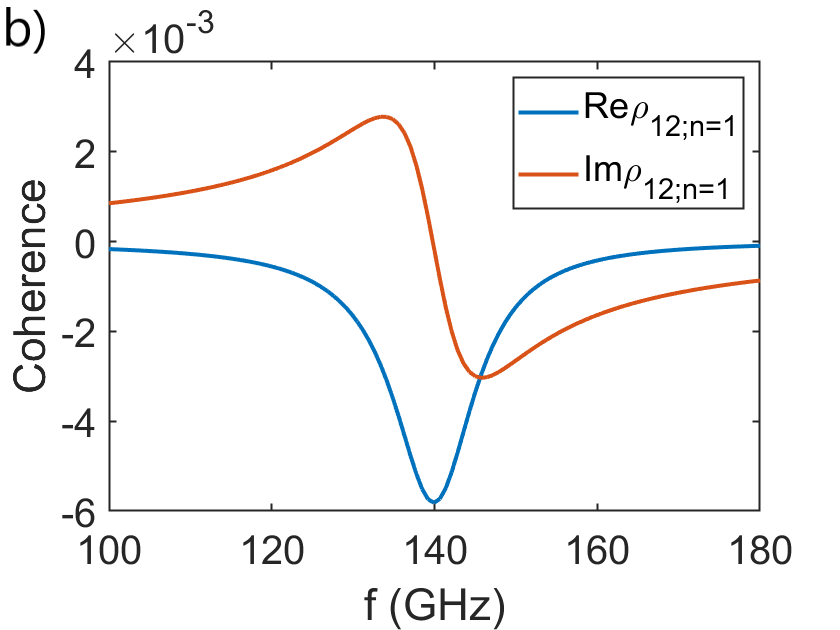}
\caption{Change in the populations a) and coherence, Floquet $n=1$, b) of the states vs frequency. The applied bias is $V_{DC}=7.7$ mV while the rest of parameters are defined in Sec.~\ref{Parameters}. The resonance peak is found at 140 GHz. The change in the populations, $\Delta \rho_{ll}(f)=\rho_{ll}^{\rm Off}-\rho_{ll}^{\rm On}(f)$, of the occupied states decrease at the resonance while the state $|3\rangle$ increases twice in comparison. Real part of the coherence, $n=1$, presents the same behaviour as the populations of the occupied states while the imaginary part has a fano shape.
The changes in coherences and in populations are in the range of 0.6\% and 0.02\% respectively, which matches the relative change in current at the Larmor frequency in Fig. \ref{I0_vs_omega}. }
\label{rhos_vs_omega}
\end{figure}

We obtain clear ESR peaks in the long-time averaged current together with a characteristic  dynamical behavior of the reduced-density matrix that allows us to unambiguously conclude that Rabi oscillations are being excited.

The long-time averaged current corresponds to the Floquet component $n=0$ of the time-dependent current.
We find it useful to analyze the following magnitude~\cite{Delgado_Rossier_pss_2017},

\begin{equation}
\Delta I_{DC}(f)=I_{DC}^{\rm Off}-I_{DC}^{\rm On}(f), 
\label{DeltaI}
\end{equation} 
where $I_{DC}^{\rm Off}$ is the current far from resonance satisfying two conditions: 
$i)$ $\omega=2\pi f$ is much larger or smaller than the Larmor frequency, and $ii)$ $f$ is much smaller than the $eV_{DC}/h$. The change in the population follows the same definition.

Figure \ref{I0_vs_omega}a shows $\Delta I_{DC} (f)$ for the frequency $f$ going from zero to 1500 GHz ($\sim 6.2$ meV)  when the applied bias is 7.7 mV. Below 500 GHz one narrow peak dominates the spectra.
Figure~\ref{I0_vs_omega}b) shows a zoom of the low-frequency peak. This is clearly an ESR peak due to the Rabi oscillations between states $|1\rangle$ and $|2\rangle$. The origin of the oscillation is a spin-flip transition that reflects the Zeeman splitting induced by the applied magnetic field, $\mathbf{B}$. The Larmor frequency due to the Zeeman splitting is then $g \mu_B|\mathbf{B}|/h\approx 5.79\times 10^{-1}/h$ meV $\approx 140$ GHz, in excellent agreement with the frequency that maximizes $\Delta I_{DC}$. 

Above  500 GHz we can identify several inelastic thresholds 
in the current: $f_1=(\Delta_{13}-\mu_{R})/h=619$ GHz, $f_2=(\Delta_{23}-\mu_{R})/h=759$ GHz, $f_3=(\Delta_{32}+\mu_{L})/h=1103$ GHz and $f_4=(\Delta_{31}+\\mu_{L})/h=1243$ GHz.
If $V_{AC}$ is set to zero, these inelastic transitions vanish. This shows that the thresholds are associated with absorption of energy quanta given by the above frequencies, $h f_i$. The high-frequency steps show a strong dependence on the temperature.  The shape of the steps is directly related to the Fermi occupation factors.  At larger frequencies all transport channels are open and the current saturates.

The hallmark of ESR transitions is revealed in the reduced-density matrix dynamical behavior. 
Figure \ref{rhos_vs_omega}a shows the populations of the states as the driving frequency is ramped up. 
Figure~\ref{rhos_vs_omega}b) shows the real and imaginary part of the coherences between states $|1\rangle$ and $|2\rangle$, that are the ones involved in the Rabi oscillation. The resonant feature is centered around the Larmor frequency. 
The behavior of the reduced-density matrix unequivocally proves that the peak in the long-time averaged current at 140 GHz is associated to an ESR. 
In other words, we have demonstrated that the modulation of the tunnel barrier by an oscillating electric field is able to induce ESR.

The resonance peak in Fig. \ref{I0_vs_omega}b) can be fitted to a Fano line shape, given by:
\begin{equation}
\Delta I_{DC}= \frac{2c}{\Gamma}\frac{1}{q^{*2}+1}\frac{[1+q^*(f-f_0)/\Gamma]^2}{1+4\left[(f-f_0)/\Gamma\right]^2},
\label{Fano_shape}
\end{equation} 
where $\Gamma$ is the width of the resonance, $f_0$ is the resonance frequency, $c$ is a parameter proportional to the square of the Rabi frequency and related to $T_1$ processes and $q^*=1/q$, being $q$ the commonly called Fano parameter. 

The motivations behind the use of a Fano line shape are the following. 
First, Fano line shapes are derived within the optical Bloch equations description of ESR~\cite{Abragam_Bleaney_book_1970}. The difference here is that we have an effective Rabi frequency induced by the time-dependent tunneling and no pure dephasing. 
Secondly, the fitting to a Fano line shape has been implemented in most of the ESR experiment \expSTMESR\  to extract parameters such as Rabi frequency or coherence times. And finally, even the simplest case $S=0$ becomes challenging to solve analytically, which makes the fitting analysis quite essential to gather information on how the resonance features change with the external parameters.
Particularly, in Eq. (\ref{Fano_shape}), the rate functions are contained in the width $\Gamma$ and $c$ parameter, as well as in the Fano parameter, such that they control the line shape. For the present case, $q^*=-0.088$, $c=5.021$ pA GHz and $\Gamma=12.1$ GHz, therefore $I_{peak}\approx 2c/\Gamma=0.82$ pA. Since $q^*$ is a small number, a Lorentzian line shape can be a good approximation to the problem as Bloch Redfield equations indicate~\cite{Delgado_Rossier_pss_2017}. From the width we can extract the decoherence time since $T_2=1/\pi\Gamma$ obtaining $T_2=0.026$ ns. This value coincides exactly with the one of Eq. (\ref{coherence}) which, for $S_T=1/2$ and the parameters used ($B_x\gg B_z$), is
\begin{equation*} 
T_2^{12}\approx \frac{\hbar}{\lambda_{31\uparrow}^* \lambda_{32\uparrow}  \gamma_{R}'}=\frac{2\hbar B(B+Bz)}{B_x^2\gamma_{R}'}=0.026\ ns.
\end{equation*}
Finally, $I_{peak}$ is proportional to the Rabi frequency of the resonance, as we will explicitly verify in the next sections. As a side note, since the width and $c$ parameter are related to the coherence time and the induced Rabi frequency, they depend on the Fermi distribution function. This implies that the resonance profile can be significantly modified by the temperature as we will see later on. 

\subsubsection{Spin polarization of the electrodes}
\label{polarization}
\begin{figure}
\centering
\includegraphics[width=0.8\linewidth]{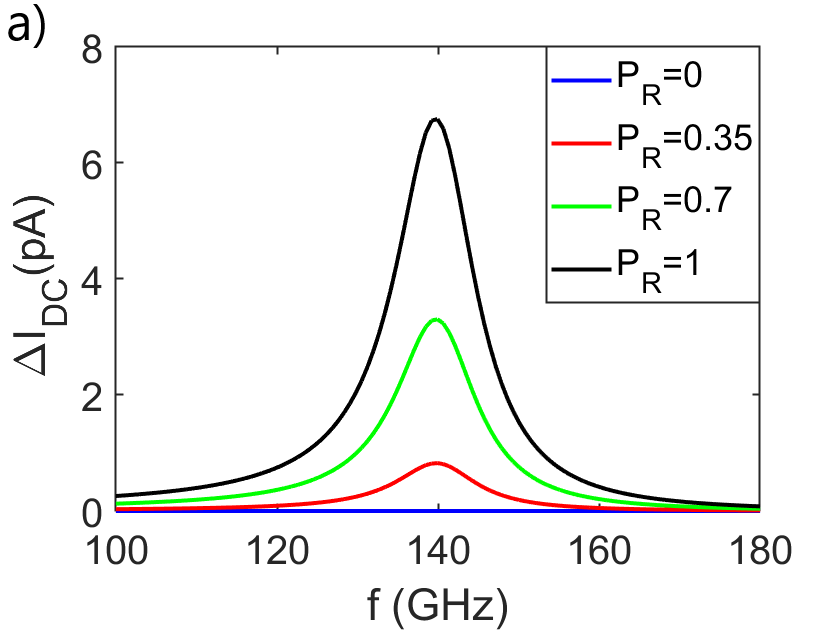}
\includegraphics[width=0.8\linewidth]{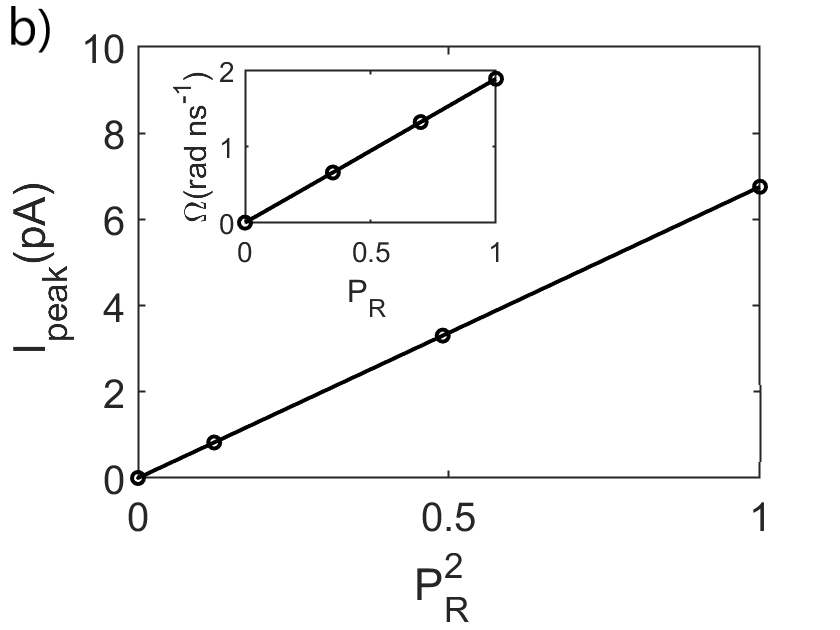}
\caption{a) Change in the current vs frequency close to the resonance fixing all the model parameters to the ones in Sec.~\ref{Parameters} but modifying the polarization. The resonance peak height grows quadratically with the polarization. This is demonstrated in panel b) where the peak height (main panel) and the Rabi frequency (inset) are plotted as a function of $P_R^2$ and $P_R$ respectively.
Negative polarization values implying the right electrode magnetic moment is antiparallel to the z component of the magnetic field it only introduces a
$\pi$-phase change in the coherences at resonance for $S_T=1/2$.
}
\label{I0_vs_omega_P}
\end{figure}

Let us study the role of the electrode polarization in the behavior of the ESR resonance. Results are shown in Fig. \ref{I0_vs_omega_P}. Qualitatively, the role of the electrode polarization can be  understood by noticing that  it selects certain spin directions for the transport. Hence  spin-flip processes are forced in order to contribute to the time-dependent current.

We can also see that a finite value of the polarization is important to get a non-zero value of the Rabi frequency. This can be verified by calculating the Rabi frequency from Eq. (\ref{rabi}), which is
\begin{equation}
\hbar \Omega_{12;1} =  A_{R} \lambda_{31\uparrow}^* \lambda_{32\uparrow}  \gamma_{R}'P_R/2,
\label{Result_Rabi_S=0}
\end{equation}
where we used $\lambda_{31\downarrow}^* \lambda_{32\downarrow}=-\lambda_{31\uparrow}^* \lambda_{32\uparrow}$. 
On the other hand, since $I_{peak}\propto P_R^2$, we conclude that $I_{peak}\propto \Omega_{12;1}^2$. This is clearly observed in Fig. \ref{I0_vs_omega_P}b) where the peak height follows a perfect linear dependence on the square of the polarization.
Interestingly, this matches the experimental observations of Ref.  \onlinecite{Willke_Paul_sciadv_2018, kim2021spin},  where 
$ I_{peak}$ is found to be  proportional to the square of the Rabi frequency, $\Omega^2$, under the constrain that $\Omega T_1 T_2\ll 1$,
with $T_1$ and $T_2$ the involved life and coherence times, respectively. 

Equation (\ref{Result_Rabi_S=0}) makes clear the key role of the electrode polarization. First, it is responsible of the finite Rabi term.
Second, the tip polarization direction determines the direction of the effective driving magnetic field inducing the Rabi oscillations. 

Interestingly, the width $\Gamma$ of the ESR resonance remains constant as the polarization changes.
Since the width is proportional to the inverse of the coherence time $1/T_{2}^{12}$, this quantity also remains essentially constant as the polarization changes. The same behavior is obtained in the framework of Bloch-Redfield theory~\cite{Delgado_Rossier_pss_2017,Willke_Bae_science_2018}.  

\subsubsection{Driving amplitude}

\begin{figure}
\centering
\includegraphics[width=0.9\linewidth]{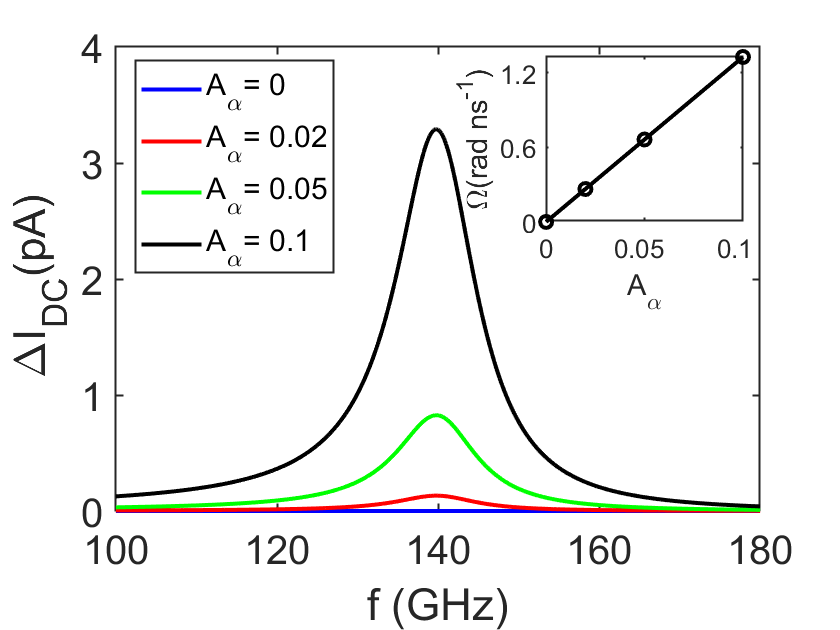}
\caption{Change in the current vs frequency for different driving amplitudes close to resonance. Others model parameters employed are found in Sec.~\ref{Parameters}. Inset shows the exact linear dependence of the Rabi vs the driving amplitude.}
\label{I0_vs_omega_A}
\end{figure}

The driving term is obviously essential to have any resonance signal. 
We found that the driving is proportional to the applied alternating bias, $V_{AC}$.
It is then instructive to study how the ESR amplitude depends on the applied bias.
Using the parameters in 
Sec. \ref{AC_driven} leads to Fig. \ref{I0_vs_omega_A}. 
Clearly, zero driving implies no resonance while increasing the driving provides a higher peak. 
One can notice in the inset of Fig. \ref{I0_vs_omega_A} that the Rabi frequency is clearly linear with the driving amplitude
that implies a quadratic dependence on the current peak. This is not a surprise  recalling the behavior of the DC component of Eq. (\ref{Current_floquet_simp}). 
The exact dependence on the driving is hidden in the Floquet density matrix elements that are connected to four Floquet numbers 
$\pm 1, 2$ as we mentioned in Sec.~\ref{sec_theory}. Moreover, we found that $ I_{peak} \propto \Omega_{12;1}^2$.

For large $V_{AC}$, the Taylor expansion leading to Eq. (\ref{hopping}) will break down and higher
harmonics of $\omega$ will contribute. This could explain  the experimentally found saturation of the ESR peak with applied bias \cite{Willke_Paul_sciadv_2018}.

\subsubsection{Dependence with temperature}

\begin{figure}
\centering
\includegraphics[width=0.8\linewidth]{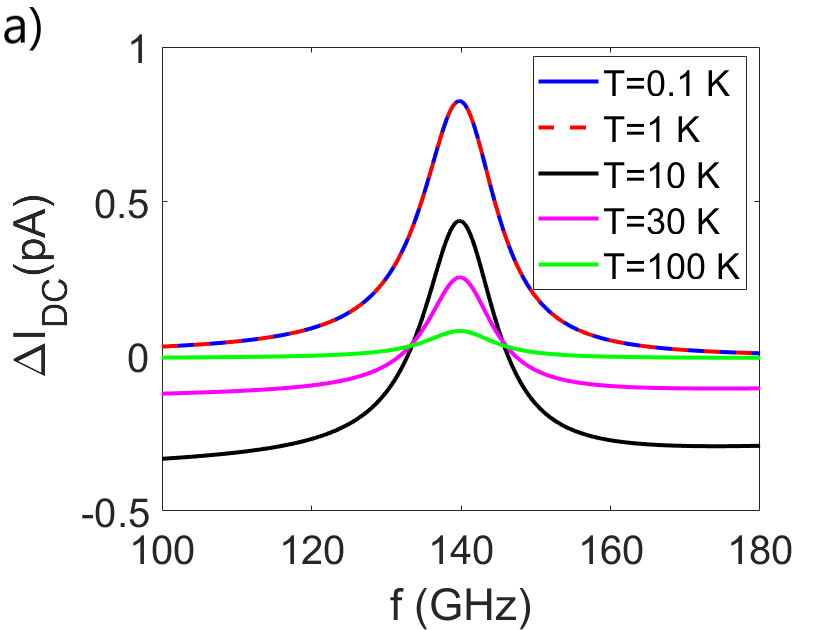}
\includegraphics[width=0.8\linewidth]{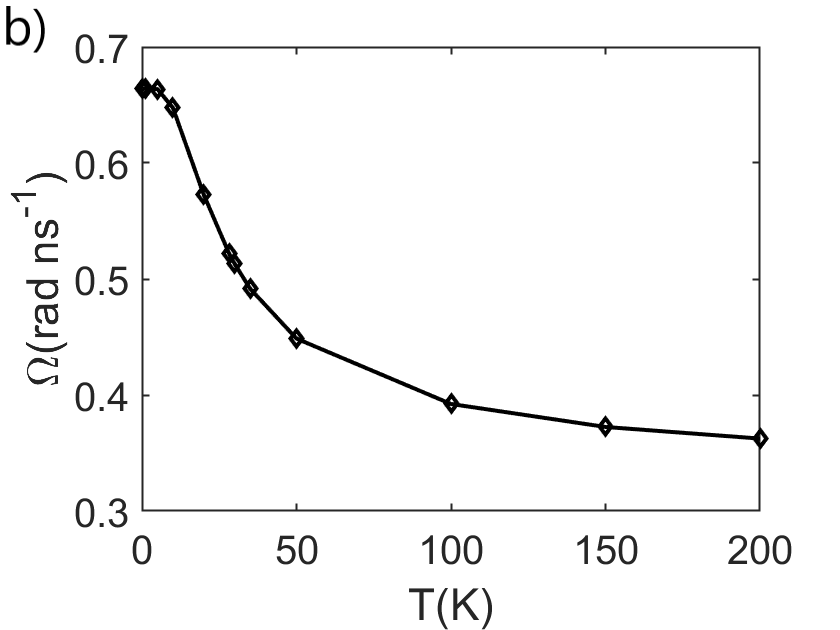}
\caption{Current change a) and Rabi frequency b) for different temperatures while the rest of the parameters are found in Sec.~\ref{Parameters}. 
The resonance peaks for the 0.1 and 1 K coincide, reflecting the fact that at the present parameters low-temperature decoherence is
produced by the applied bias.  As the temperature increases, the ESR becomes wider and shorter as a consequence of the
enhanced decoherence.
Surprisingly, the Rabi frequency also decreases with temperature, showing that temperature also has an influence on the
coupling between resonating states beyond the effects of decoherence.
The ESR peak height scales with the square of the Rabi frequency, such that an additional reduction in the Rabi frequency translates into a considerable reduction of the resonance peak.
}
\label{Cosas_T}
\end{figure}

The main temperature effect on the ESR at large DC voltage is to enhance the decoherence rate, $\hbar/T_2^{12}$,
that leads to wider and smaller ESR peaks. However, increasing the temperature leads to additional changes in the $\Delta I_{DC}$ resonance profile. This is shown in Fig. \ref{Cosas_T}a). 
Interestingly,
the reduction of the ESR height is not only due to the increase of the resonance width. Indeed, we
find a reduction of the Rabi frequency as the temperature increases, Fig. \ref{Cosas_T}b). 
The temperature dependence of the Rabi frequency implies an intrinsic reduction of the efficiency of the spin-flip
process to produce the ESR.  

Between $0.1$ K and $1$ K we are in the condition that all electronic channels are open,
$|eV_{DC}|\gg k_BT,\; \Delta_{21}$, the non-adiabatic decoherence (also known as population scattering) is controlled by the bias voltage, and not by the thermal broadening \cite{Delgado_Rossier_pss_2017}. 
As  seen in Fig. \ref{Cosas_T}) and b),  there is not evolution in the small temperature range.

As the temperature increases, it starts to dominate the behavior of the resonance, decreasing the ESR peak amplitude, hence becoming barely observable for $k_B T\gg |eV_{DC}|,\; \Delta_{21}$. 
However, the Rabi frequency tends to saturation, Fig. \ref{Cosas_T}b), showing that ESR is active despite being barely observable.

Another source of temperature dependence that we have not considered here is the variation of the tip-polarization with the temperature \cite{Yang_Bae_prl_2017,Y_Bae_advanced_science_2018}. This could be easily implemented to compare with the experimental results at various temperatures.

\subsubsection{Magnetic-field dependence}

\begin{figure}
\centering
\includegraphics[width=0.8\linewidth]{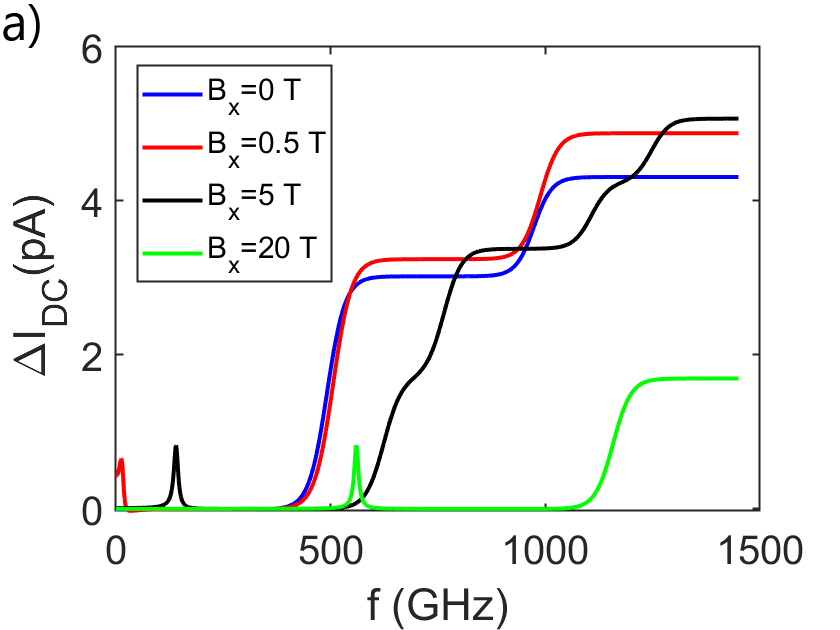}
\includegraphics[width=0.8\linewidth]{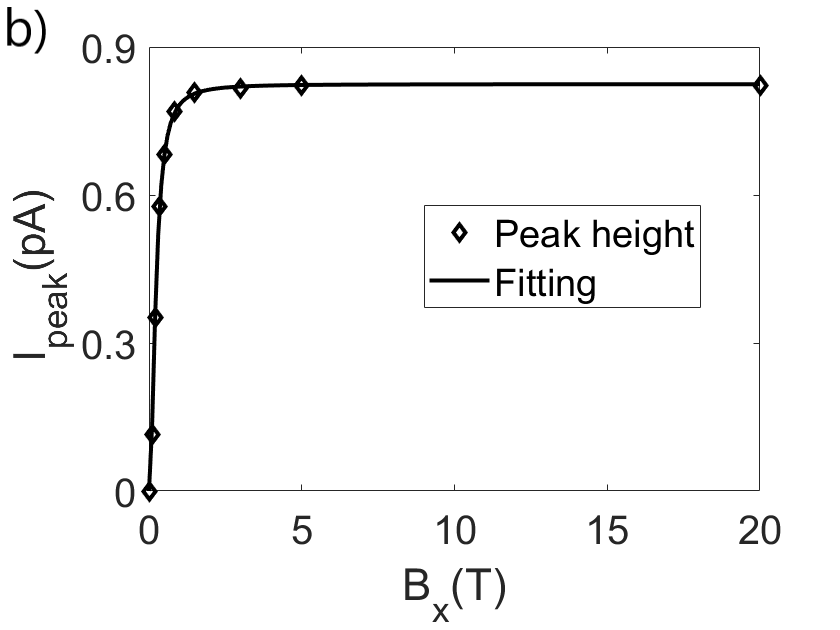}
\includegraphics[width=0.8\linewidth]{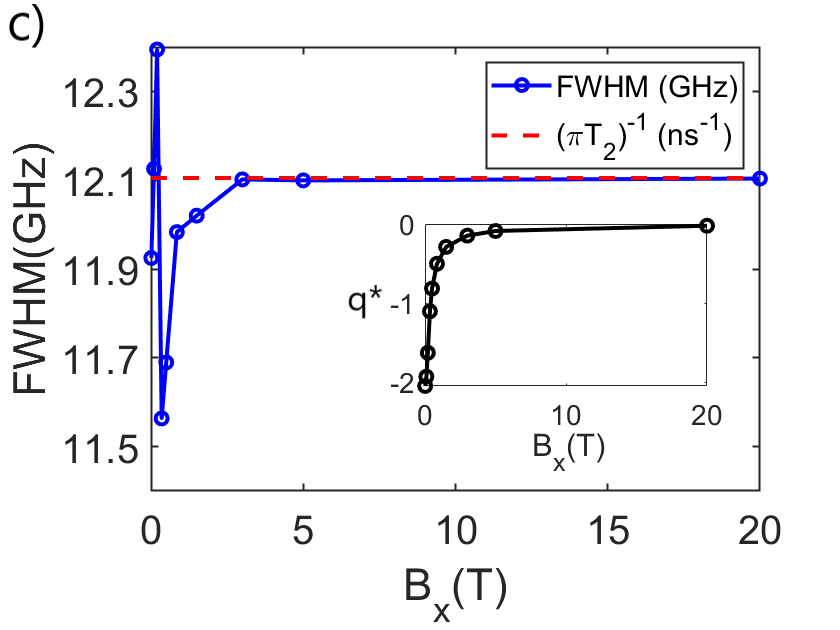}
\caption{a) Change in the long-time averaged current \textit{vs} driving frequency for different values
of the $x$-component of the magnetic field, $B_x$, while keeping the
$z$-component fixed at 0.2 T. Section \ref{Parameters} contain the rest of the model parameters used.  The spectra shift linearly with $B_x$ following
the Zeeeman shift. Zero $B_x$ field leads to no resonance, and increasing it provides a needed mixture of states $|1\rangle$ and $|2\rangle$ that rapidly saturates. The resonance peak positions from lower to higher are: 15.07, 140.06, 559.84 GHz. In b) the resonance
peak height is plotted as a function of $B_x$. The saturation of
state mixing is reached at 2 T and as a consequence, the peak height also saturates. The fitting curve is $aB_x^2/(B_x^2+b)$ where $a = 0.82\ pA,\ b = 0.053\ T^2 \sim B_z^2$ showing that it is proportional to the square of the Rabi frequency, Eq. (\ref{RabiB}).
Panel
c) shows the inverse of the Fano parameter, $q^*$ (inset), and the resonance FWHM 
as a function of $B_x$. These data show that the line profile becomes closer to a Lorentzian profile
as the magnetic field increases because the Larmor frequency fixed by the Zeeman energy increases, placing
the resonance away from zero frequency.
}
\label{I0_vs_omega_B}
\end{figure}

In the case of  spin-$1/2$ systems, the perpendicular magnetic field $B_x$ is essential to have any sort of coherence since it allows having states $|1\rangle$ and $|2\rangle$ with mixed character, which is necessary to have an ESR-active system~\cite{Baumann_Paul_science_2015,J_Reina_Galvez_2019}. 
In order to illustrate the role of $B_x$, we fix all the parameters except $B_x$ to those introduced in Sec. \ref{AC_driven}. Figure \ref{I0_vs_omega_B}a) shows the results of $\Delta I_{DC}$ as a function of the driving frequency $f$ for different values of $B_x$. At zero transverse field, no resonance peak appears. As $B_x$ increases, a resonance peak develops in the frequency position, depending on the total magnetic field. The peak height is related to the strength of the state mixing~\cite{Baumann_Paul_science_2015,J_Reina_Galvez_2019}. We have shown in the polarization section \ref{polarization} that for the three-level case of $S_T=1/2$, the Rabi frequency is proportional to $\lambda_{31\sigma}^*\lambda_{32\sigma}$. The eigenvectors of the impurity Hamiltonian are:
\begin{eqnarray}
|1\rangle&=&\frac{1}{\sqrt{2B}}\frac{1}{\sqrt{B+ B_z}}[-(B+B_z) |\downarrow\rangle+B_x|\uparrow\rangle], \crcr
|2\rangle&=&\frac{1}{\sqrt{2B}}\frac{1}{\sqrt{B- B_z}}[(B_z-B)|\downarrow\rangle-B_x|\uparrow\rangle],\crcr
|3\rangle&=&|0\rangle. \nonumber
\end{eqnarray}
Hence, $|3\rangle$ corresponds to an empty electronic level. Here, $B=\sqrt{B_z^2+B_x^2}$. 
These results show that only if $B_x \neq 0$, $|1\rangle$ and $|2\rangle$ contain both spin up and down components, enabling Rabi oscillations~\footnote{The limit $B_x \rightarrow 0$ has to be taken with care, but confirms the above explanation.}. 

Using these eigenstates, we can calculate the projections entering the Rabi frequency expression. We get 
\begin{eqnarray}
 \hbar \Omega_{12;1} &=& A_R \sum_\sigma \lambda_{31\sigma}^*\lambda_{32\sigma}(1/2+\sigma P_R)\gamma_{R}' \nonumber \\
&=& A_R  P_R\gamma_{R}'B_x/2B
\propto \sqrt{\frac{B_x^2}{B_x^2+B_z^2}},
\label{RabiB}
\end{eqnarray}
that tends to a constant value as $B_x$ increases. 
This is shown in Fig. \ref{I0_vs_omega_B}b). Notice that in this case, for values $B_x > 2$ T saturation is attained,
reflecting that the mixing of spin up and down states cannot be further enhanced by the $x$-component of the  magnetic field
for a fixed valued of $B_z=0.2$ T. 

Furthermore, the magnetic field can affect the Fano parameter and the width of the resonance. 
The inset of Fig. \ref{I0_vs_omega_B}c) shows the $q^*$ dependence on the $x$-component of the  magnetic field. 
Only at very low magnetic fields $|q^*|$ increases drastically. This is due to the distorted resonance profile 
close to zero frequency, which is also reflected in the values of the full width at half maximum (FWHM), 
Fig. \ref{I0_vs_omega_B}c). As the values of $B_x$
increase, the ESR moves away from zero freaquency and
develops a full-fledged Fano profile.  The FWHM becomes constant then, since it is proportional to 
the inverse of the coherence time, $1/T_2^{12}$.

\section{Results for two coupled spins $S_T=0,1$}

In this section we illustrate the workings of the tunneling-induced ESR for two interacting spin-1/2 systems,
where one of them is coupled to two electrodes. Thus, we consider a localized $S=1/2$ spin 
exchange coupled to the electronic level that transfers electrons between the two biased electrodes. Hence, the electronic level acts as an extra spin-1/2 system when it is singly occupied. 
The basis set of the impurity is given by the tensor product of the three states of the electronic level (down, up and empty) times the two states of the coupled spin-1/2 (down and up), a total of six states.

\begin{figure}
\centering
\includegraphics[width=1.\linewidth]{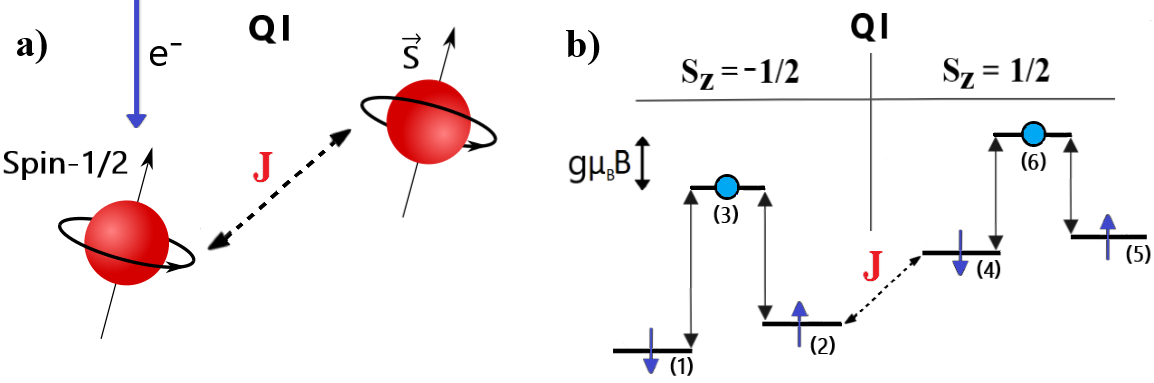}
\caption{Scheme for the $S=1/2$ case coupling with the incoming electron from the electrodes. Panel a) shows both systems
with an exchange-coupling strength $\it J$. Three triplet states ($S_T=1$) form plus a singlet ($S_T=0$).Panel b) shows a simplification of the energy-level scheme similar to Fig. \ref{Scheme_S=0}. The scheme depicts a system with negative $\varepsilon_\sigma$
and a  magnetic field parallel to the $z$ direction. Due to exchange interaction, both spins interact and the  $x$ or $y$ components of the
magnetic field can mix the filled  and empty states. }
\label{Scheme_S=0.5}
\end{figure}

Figure \ref{Scheme_S=0.5} shows a  sketch of the system. In the coupled case, assuming a ferromagnetic coupling $(\it J<$ 0) and $|\it J|$ larger than the Zeeman slitting
the full system becomes a $S_T=1$ spin. The triplet states are energetically below the singlet state,
leading to an effective spin-1 situation. Contrary, if the coupling is anti-ferromagnetic ($\it J>$ 0), the singlet state is the one with the lowest energy and the spin of the system is $S_T=0$. When $\it J$ is close to zero, we have a decoupled $S_T=1/2$ spin and the corresponding results can be found in the previous section.

Hence, this model allows us to explore the physics of ESR in a magnetic impurity of spin  $S_T=1$, or of two impurities of spin 1/2. In particular, this last case has been carefully analyzed in a series of STM-ESR experiments on two Ti atoms, at different
distances on a MgO/Ag (100) substrate~\cite{Yang_Bae_prl_2017,Y_Bae_advanced_science_2018}. The distance between the two Ti atoms
control the exchange coupling strength $J$, and the bias between tip and sample contains an alternating component. The measured
DC tunneling current shows four distinct peaks at different driving frequencies that have been associated with
singlet-triplet transitions under an external magnetic field and the local field of the
magnetic tip~\cite{Yang_Bae_prl_2017,Y_Bae_advanced_science_2018}.

If $J$ and the transverse magnetic field are non zero, the filled and empty states are mixed, making  possible the appearance of resonances between states with $ \Delta S_{T,z}  = S_{T,z_i}-S_{T,z_j}  = \pm 1$ (spin flip), where $i$ and $j$ can be any state, labeled by 1 to 6 in Fig. \ref{Scheme_S=0.5}. For the sake of simplicity, we set $D=0$ eV in Eq. (\ref{eq:himp0}).

Similar to the $S_T=1/2$ case analyzed in the previous section, the current is computed on the left electrode in a situation of large DC bias voltage where all the states are properly connected through the sequential tunneling. The other parameters are: $i)$ Driving amplitude $A_\alpha=0.5$, $ii)$ polarization $P_R=0.45$, $iii)$ Temperature $T=1$K. 
The parameters describing the hybridization with the reservoirs are
$iv)$ Coupling constants $\gamma_{R}'=0.5\ \mu$eV  and $\gamma_{L}'=0.5$ meV that allow us to reproduce an overall long-time averaged
current of 50 pA. 
Truncating the  Floquet components at $n=5$ is enough to get a converged  solution of the time-dependent master equation.

In order to take into account the local magnetic field of the tip, we
add a new Zeeman term to the impurity Hamiltonian that only affects
localized electrons. The origin of this field in the experimental setup is the magnetic interaction exerted by the tip on the atom underneath. We label this field as $\mathbf{B_{tip}}$ following the notation of Ref. [\onlinecite{Y_Bae_advanced_science_2018}]. 

\begin{figure}
\centering 
\includegraphics[width=0.8\linewidth]{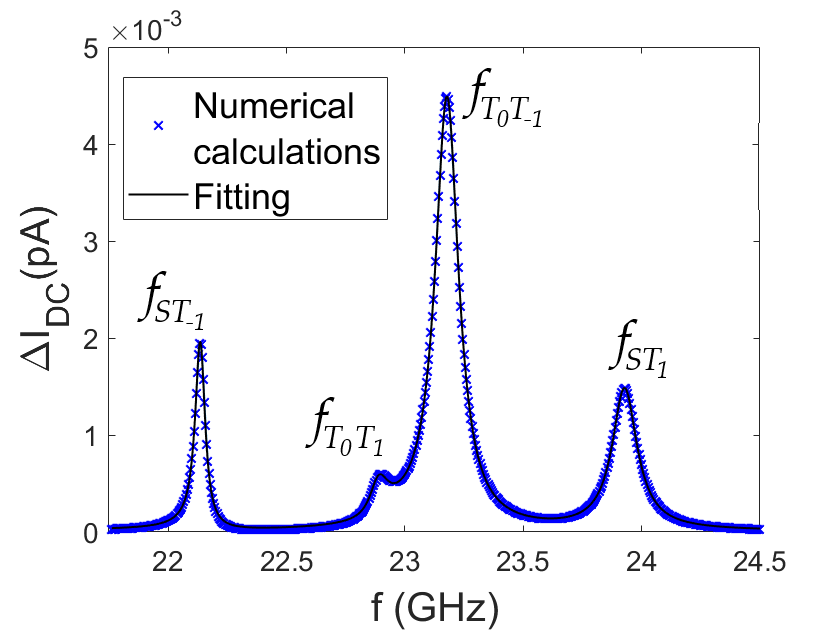}
\caption{Frequency spectrum for $S=1/2$ when an additional $\mathbf{B_{tip}}$ is applied. We used $\varepsilon_\sigma=-1$ meV, $\mathbf{B}=(0,0,0.89)$ T, $\mathbf{B_{tip}}=(0.035,0,0.05)$ T , $\it J=3.1$ $\mu$eV$=0.75$ GHz, $\gamma_R'=0.5$ $\mu$eV, $\gamma_L'=0.5$ meV, $A_\alpha=0.5$, $P_R=0.45$. In addition, we took into account the difference in magnetic moment as in the experiment of Ref. \cite{Y_Bae_advanced_science_2018}. The simulation is fitted to the sum of four Fano lineshape of the form of Eq. (\ref{Fano_shape}) and the spin flip transitions are labeled in the plot. $S$ indicates a singlet state while $T$ a triplet one.}
\label{B_tip_spectrum}
\end{figure}

Figure \ref{B_tip_spectrum} shows the result of our calculation. Due to the presence of two different magnetic fields, we find four resonance peaks. The values are: 0.051, 0.087, 0.12, 0.13 GHz respectively. The order of the resonance peaks is, from low to high frequencies: transition singlet to the triplet state $S_{T,z}=-1$,  transitions between the triplet states $ S_{T,z}=0$ and $S_{T,z}=\pm 1$ and finally singlet to triplet state $ S_{T,z}=1$ transition, in good agreement with the experimental interpretation~\cite{Y_Bae_advanced_science_2018}.

The frequency spectra can be fitted to Fano lineshapes as in the previous section. This allows us to characterize each resonance peak identifying the asymmetry through the Fano parameter, plus the peak height and width. These parameters can be associated to the Rabi frequency, the coherence and life times of each transition. 
 The Rabi frequency controls the peak height and the coherence time, $T_2$, the width, similar to the $S_T=1/2$ case. Every Rabi frequency is proportional to the polarization, but now the strength of the peak does not depend on it linearly. As we increase the (positive) polarization, the right electrode  accepts more spin-up than spin-down electrons. 
Therefore, only the transitions from the triplet state with $S_z=-1$ to any of the two  $S_z=0$ states are possible,
 since they involve flipping the spin of the incoming electron from down to up. The frequency spectra will then have two peaks: the first and the third transitions in Fig. \ref{B_tip_spectrum}.

From the fitting analysis, we notice an increase of the width of the peaks as  the driving frequency increases. 
 This happens because we are in an asymmetric coupling case where $\gamma_L'\gg \gamma_R'$. It also 
 generates  more asymmetric Fano lineshapes. The first and third resonances have negative and similar inverse Fano factors, $q^*$,
equal to -0.07 and -0.03 respectively, while the other two transitions have positive values equal to 0.2 and 0.007 for the second and fourth transition.

%% file: Appendices.tex
\section{Derivation of the Floquet master equation}
\label{derivation_floquet_master_equation}

Here we summarize the steps to derive the equation describing the dynamics of the reduced density matrix following  
Ref. \cite{Bibek_Bhandari_2021_nonequilibrium}. 
First, we take into account that expectation value at a given time $t$ of
any operator ${\cal O}$ acting on the Hilbert space of the impurity can be written as
\begin{equation} \label{ot}
\langle {\cal O} \rangle (t) = \mbox{Tr}\left[\hat{\rho}_{\rm T} (t) {\cal O}(t) \right] = \sum_{lj} O_{lj}(t) \rho_{lj}(t),
\end{equation}
where $O_{lj}=\langle l | {\cal O} |j \rangle $ are the matrix elements of the operator ${\cal O}$ in the basis of eigenstates of $H_S$. 
Here, we have introduced the definition of the elements of the reduced density matrix,
\begin{equation*}
\rho_{lj} (t) = \mbox{Tr}\left[\hat{\rho}_{T} (t)  \hat{\rho}_{lj} \right],
\end{equation*}
with the trace taken over the degrees of freedom of the total system including the impurity, the reservoirs and the couplings between them. In the present problem $\hat\rho_{lj}=|l\rangle\langle j|$ resemble the so-called Hubbard operators~\cite{Hewson_book_1997} and are defined for any pair $|l\rangle, \; |j\rangle$ of many-body eigenstates  of the Hamiltonian $H_{\rm S}$. Our goal is the derivation of the equation ruling the dynamics of $\rho_{lj}(t)$. Changing to the Heisenberg picture with respect to the full Hamiltonian (denoted with “$H$”), we can write
$\rho_{lj}(t)={\rm Tr}(\hat{\rho}_{0}\hat{\rho}_{lj}^{H}) $ with $\hat{\rho}_{lj}^H(t)= U(t,t_0)\hat{\rho}_{lj} U^{\dagger}(t,t_0)$ being $\hat{\rho}_{0}$ the density matrix at the initial time $t_0$ and $U(t,t_0)=\hat{T}\left[e^{-i\int_{t_0}^t dt'H(t')/\hbar} \right] $ the evolution operator with $\hat{T}$ the time-ordering operator. 
 Applying the Heisenberg equation of motion leads to

\begin{eqnarray}
&&\frac{d\rho_{lj}(t)}{dt}=\frac{d \langle \hat{\rho}_{lj}^{H}\rangle }{dt}=\frac{i}{\hbar}\left\langle \left[\hat{\cal H}^{'H}_{S},\hat{\rho}^H_{lj}\right]\right\rangle \nonumber\\ 
\pm \frac{i}{\hbar}\sum_{\alpha k \sigma u}&&\left[ w_{\alpha }\lambda_{ul\sigma}\langle c^{\dagger H}_{\alpha k\sigma }\hat{\rho}_{uj}^H\rangle - w_{\alpha }\lambda_{ju\sigma}\langle c^{\dagger H}_{\alpha k\sigma }\hat{\rho}^H_{lu}\rangle\right.   \nonumber\\
&&+\left.w^{*}_{\alpha }\mu_{ul\sigma} \langle \hat{\rho}^H_{uj}c_{\alpha k\sigma }^H \rangle  - w^{*}_{\alpha }\mu_{ju\sigma} \langle  \hat{\rho}_{lu}^H c_{\alpha k\sigma }^H \rangle  \right],\nonumber\\ 
\label{rho_master_eq_A}
\end{eqnarray}
where we have use the fact that $\hat{\rho}_{uv}\hat{\rho}_{lj}=\hat{\rho}_{uj}\delta_{vl}$. Omitting all the superindexes “$H$”, the terms that combine the $c_{\alpha k \sigma}\ (c_{\alpha k \sigma}^{\dagger})$ and $\hat{\rho}_{lj}$ invite us to define Green's function such as:
\begin{eqnarray}
G^{<}_{lj,\alpha k \sigma}(t,t')&=& \pm i\langle c^{\dagger}_{\alpha k \sigma}(t') \hat{\rho}_{lj}(t) \rangle  \nonumber \\
G^{<}_{\alpha k \sigma,lj}(t,t')&=& \pm i\langle \hat{\rho}_{jl}^{\dagger}(t')c_{\alpha k \sigma}(t)\rangle.
\label{Mixed_Green}
\end{eqnarray}
The upper/lower signs apply to many-body states $|l\rangle, \; |j\rangle$ of the impurity that differ in an odd/even number of fermions (we recall that the number of particles is conserved in $H_{\rm S}$). To implement perturbation theory in the couplings between the impurity and the reservoirs, $w_{\alpha }^0$,
relevant to the case where this parameter is the lowest energy scale of the full Hamiltonian $H$, it is convenient to define the interaction picture with respect to the uncoupled Hamiltonian $h=H_{\rm res}+H_{\rm S}$. 
The mix Green's function is then calculated by recourse to Langreth theorem in the Schwinger-Keldysh contour, by considering
the initial time $t_0\rightarrow -\infty$ and the initial state corresponding to $h$ \cite{rammer_2007}. 
At the linear order of perturbation theory
\begin{eqnarray}
&&G_{lj,\alpha k \sigma}^{<}(t,t')\approx \frac{1}{\hbar}\int_{-\infty}^{+\infty} dt_1\sum_{uv}\lambda^*_{vu\sigma}w^{*}_{\alpha }(t_1)\times
\nonumber \\
&&\left[g_{lj,uv}^{r}(t,t_1)g_{\alpha k\sigma}^{<}(t_1-t')+ g_{lj,uv}^{<}(t,t_1)g_{\alpha k\sigma}^{a}(t_1-t') \right]\nonumber \\
&&
\label{mix_green_1}
\end{eqnarray}
and
\begin{eqnarray}
&&G_{\alpha k \sigma,lj}^{<}(t,t')\approx \frac{1}{\hbar}\int_{-\infty}^{+\infty} dt_1\sum_{uv}\lambda_{uv\sigma}w_{\alpha }(t_1)\times
\nonumber \\
&&\left[g_{\alpha k\sigma}^{r}(t-t_1)g_{uv,lj}^{<}(t_1,t')+ g_{\alpha k\sigma}^{<}(t-t_1) g_{uv,lj}^{a}(t_1,t')\right],\nonumber \\
&&
\label{mix_green_2}
\end{eqnarray}

The lesser Green's functions for these isolated systems that enter these expressions are 
\begin{eqnarray}
g^{<}_{lj,uv}(t,t')&=& \pm i\langle \hat{\rho}^{h \dagger}_{vu}(t') \hat{\rho}^h_{lj}(t) \rangle \nonumber\\
& =& i \langle \hat{\rho}_{uj}(0)\rangle \delta_{lv} e^{-\frac{i}{\hbar}(E_u-E_v)(t-t')} e^{iEt/\hbar},\nonumber \\
g^{<}_{\alpha k \sigma }(t,t')&=& i\langle c^{\dagger h}_{\alpha k \sigma}(t') c^h_{\alpha k \sigma}(t)\rangle,
\end{eqnarray}
with $E=E_l-E_j+E_u-E_v  $ and being $E_{i}$  the eigenenergies of the Hamiltonian of the decoupled impurity.
The corresponding greater functions read $g^{>}_{\nu,\nu^{\prime}}(t,t') = \left[ g^{<}_{\nu^{\prime},\nu}(t',t)\right]^*$, while the retarded and advanced ones are
$g^r_{\nu,\nu^{\prime}}(t,t')= \theta(t-t')\left[ g^{>}_{\nu,\nu^{\prime}}(t,t')- g^{<}_{\nu,\nu^{\prime}}(t,t')\right]$ and $g^a_{\nu,\nu^{\prime}}(t,t')=\left[ g^{r}_{\nu^{\prime},\nu}(t',t)\right]^*$, respectively. The latter can be expressed as
\begin{eqnarray*}
g^{r}_{lj,uv}(t,t')&=&
-i\theta(t-t')e^{-\frac{i}{\hbar}(E_u-E_v)(t-t')} \nonumber \\ 
&&\times(\langle\hat{\rho}_{uj}(0)\rangle \delta_{vl}+\langle\hat{\rho}_{lv}(0)\rangle \delta_{ju})e^{iEt/\hbar},
\end{eqnarray*}
The Fourier transform of the Green's functions of the electrodes are $g^<_{\alpha k\sigma}(\epsilon)= 2\pi i f_{\alpha} (\varepsilon_{\alpha k \sigma})\delta (\epsilon - \varepsilon_{\alpha k \sigma})$,
and 
$g^{r}_{\alpha k\sigma}(\epsilon)= [\epsilon+ i0^+ -\varepsilon_{\alpha k \sigma}]^{-1}$. 
We now define self-energies
\begin{equation}
\Sigma_{\alpha \sigma}^{c}(t,t')=\frac{1}{\hbar}\sum_{ k}w^*_{\alpha }(t)g_{\alpha k \sigma}^{c}(t-t')w_{\alpha  }(t'),
\label{selfenergy_t}
\end{equation}
with $c\equiv r,a,>,<$. Taking into account the periodic time-dependence of the tunneling parameters, Eq. (\ref{selfenergy_t}) can be expressed in terms of Floquet-Fourier components as  
\begin{equation}\label{Fourier-Floquet_Self-energy}
\Sigma_{\alpha \sigma}^{c}(t,t')= \sum_{n,n^{\prime}}
\int_{-\infty}^{+\infty} \frac{d\varepsilon}{2\pi\hbar} e^{-i(\varepsilon+n\omega) t +i(\varepsilon+n'\omega)t'} \Sigma_{\alpha\sigma,nn'}^{c}(\varepsilon),
\end{equation}
with
\begin{eqnarray*}
& & \Sigma_{\alpha\sigma,nn'}^{c}(\varepsilon)=\frac{1}{\hbar} |w_{\alpha }^0|^2\left[\delta_{n,0} +\frac{A_{\alpha}}{2}\left(\delta_{n,-1}+\delta_{n,1}\right) \right] \nonumber \\
&&\qquad\times \sum_{k} g^c_{\alpha k\sigma}(\hbar\varepsilon) \left[\delta_{n',0} +\frac{A_{\alpha}}{2}\left(\delta_{n',-1}+\delta_{n',1}\right) \right].
\label{Fourier-Floquet_Self-energy-0}
\end{eqnarray*}
Hereafter, we consider a wide-band model with a constant density of states within the spectral window $-W \leq \varepsilon \leq W$, with $W \rightarrow \infty$ which would imply that $\sum_k g^<_{\alpha k\sigma}(\epsilon)= 2\pi i f_{\alpha} (\epsilon)\rho_{\alpha\sigma}$ and $\sum_k g^{r}_{\alpha k\sigma}(\epsilon)= - i\pi\rho_{\alpha\sigma}$. 
For convenience, we also introduce the definitions
\begin{eqnarray}
&&\Lambda_{lj,\alpha\sigma}(t)=  \int_{-\infty}^{+\infty} dt_1\sum_{uv}\mu_{uv\sigma}\times \nonumber
\\ &&\left[g_{lj,uv}^{r}(t,t_1)\Sigma^{<}_{\alpha\sigma}(t_1,t)+g_{lj,uv}^{<}(t,t_1)\Sigma_{\alpha \sigma}^{a}(t_1,t) \right] 
\label{Lambda_1}
\end{eqnarray}
\begin{eqnarray}
&&\bar{\Lambda}_{\alpha\sigma,lj}(t)= \int_{-\infty}^{+\infty} dt_1\sum_{uv}\lambda_{uv\sigma}\times\nonumber \\
&& \left[\Sigma^{r}_{\alpha\sigma}(t,t_1)g_{uv,lj}^{<}(t_1,t)+\Sigma_{\alpha \sigma}^{<}(t,t_1)g_{uv,lj}^{a}(t_1,t) \right],
\label{Lambda_hat_1}
\end{eqnarray}
so that we can rewrite the master equation as follows:
\begin{eqnarray}
\dot{\rho}_{lj}(t)&=&   \frac{i}{\hbar}(E_l-E_j)\rho_{lj}(t)+\frac{1}{\hbar}\sum_{\alpha\sigma u}\left[ \lambda_{ul\sigma}\Lambda_{uj,\alpha \sigma }(t)- \right.\nonumber\\
&&\left.\lambda_{ju\sigma}\Lambda_{lu,\alpha\sigma }(t)+ \lambda^*_{lu\sigma} \bar{\Lambda}_{\alpha\sigma,uj}(t)  - \lambda^*_{uj\sigma} \bar{\Lambda}_{\alpha\sigma,lu}(t)\right] ,\nonumber \\
&&
\label{rho_master_eq_3}
\end{eqnarray}
At this point it is interesting to mention that the assumption of small $w_{\alpha }^0$ in Eqs. (\ref{mix_green_1}) and (\ref{mix_green_2}) is usually referred to as "Born" approximation in the 
derivation of the master equation based on the evolution of the density operator \cite{Breuer_Petruccione_book_2002}, while considering the 
initial time $t_0 \rightarrow -\infty$ in these equations is equivalent to considering that the time-correlations of the bath decay fast in $t-t_1$
(this is guaranteed by the wide-band model), which is the 
usual assumption in the Markov approximation. Therefore, the underlying assumptions in the derivation of Eq. (\ref{rho_master_eq_3}) are equivalent to the
so called Born-Markov approximation. 

Introducing the representation of Eq. (\ref{Fourier-Floquet_Self-energy}) in
Eqs. (\ref{Lambda_1}) and (\ref{Lambda_hat_1}) leads to
\begin{eqnarray}
&&\Lambda_{lj,\alpha\sigma}(t)=  \frac{1}{2\pi}\sum_{nn'}e^{i(n'-n)\omega t} \sum_{uv}\lambda^*_{vu\sigma}\times \nonumber \\
&& \left[(\rho_{lv}(t) \delta_{ju}+ \rho_{uj}(t) \delta_{vl})\int_{-\infty}^{+\infty}d\varepsilon  \frac{\Sigma_{\alpha\sigma,nn'}^{<}(\varepsilon-n\omega)}{\varepsilon-\Delta_{uv}/\hbar+i0^+}+\right.\nonumber\\
&&\left.2\pi i\rho_{uj}(t) \delta_{vl}\int_{-\infty}^{+\infty}d\varepsilon \delta\left(\varepsilon-\frac{\Delta_{uv}}{\hbar}\right)\Sigma_{\alpha\sigma,nn'}^{a}(\varepsilon-n\omega) \right] ,\nonumber \\
&&
\label{Lambda_1_Floquet}
\end{eqnarray}
and
\begin{eqnarray}
&&\bar{\Lambda}_{\alpha\sigma,lj}(t)=\frac{1}{2\pi} \sum_{nn'}e^{-i(n-n')\omega t}\sum_{uv}  \lambda_{uv\sigma} \times \nonumber \\
&&\left[(\rho_{uj}(t) \delta_{vl}+\rho_{lv}(t) \delta_{ju})\int_{-\infty}^{+\infty}d\varepsilon\frac{\Sigma_{\alpha\sigma,nn'}^{<}(\varepsilon-n'\omega)}{ \varepsilon-\Delta_{vu}/\hbar-i0^+}\right.  +\nonumber \\
&& \left.2\pi i \rho_{lv}(t) \delta_{ju} \int_{-\infty}^{+\infty}d\varepsilon \Sigma_{\alpha\sigma,nn'}^{r}(\varepsilon-n'\omega)\delta\left(\varepsilon-\frac{\Delta_{vu}}{\hbar}\right)\right].\nonumber \\
&&
\label{Lambda_hat_1_Floquet}
\end{eqnarray}
With Eqs. (\ref{Lambda_1_Floquet}) and (\ref{Lambda_hat_1_Floquet}) we are able to write Eq. (\ref{rho_master_eq_3}) as Eq. (\ref{rho_master_eq_6_0}) 
by introducing the following definition for the rates
\begin{equation}
    \Gamma_{vl,ju}(t) =\sum_{n} e^{-i n^{\prime} \omega t } \Gamma_{vl,ju;n^{\prime}}(\omega)
\end{equation}
with $\Gamma_{vl,ju;n^{\prime}}(\omega)$ and $\bar{\Gamma}_{vl,ju;n^{\prime}}(\omega)$ being
\begin{widetext}
\begin{equation}
\begin{split}
 &\Gamma_{vl,ju;n^{\prime}}(\omega)=\frac{1}{2\pi} \sum_{n\alpha\sigma}\lambda_{vl\sigma}\lambda^*_{uj\sigma}\int_{-\infty}^{+\infty}d\varepsilon  \frac{1}{\varepsilon-\Delta_{ju}/\hbar+ i0^{+}}\Sigma_{\alpha\sigma,nn-n'}^{<}(\varepsilon-n\omega)   \\
&+\sum_{n\alpha\sigma}\lambda^*_{lv\sigma} \lambda_{ju\sigma} \int_{-\infty}^{+\infty}d\varepsilon \left[2\pi i\delta (\varepsilon- \Delta_{uj}/\hbar)\Sigma_{\alpha\sigma,n+n'n}^{r}(\varepsilon-n\omega)  + \frac{\Sigma_{\alpha\sigma,n+n'n}^{<}(\varepsilon-n\omega)}{\varepsilon-\Delta_{uj}/\hbar- i0^{+}} \right] ,
\label{Rate_Floquet}
\end{split}
\end{equation}
\begin{equation}
\begin{split}
&\bar{\Gamma}_{vl,ju;n^{\prime}}(\omega)=\frac{1}{2\pi}\sum_{n\alpha\sigma} \lambda^*_{vl\sigma}\lambda_{uj\sigma} \int_{-\infty}^{+\infty}d\varepsilon  \Sigma_{\alpha\sigma,n+n' n}^{<}(\varepsilon-n\omega)\frac{1}{\varepsilon-\Delta_{ju}/\hbar- i0^{+}} \\
&+ \sum_{n\alpha\sigma}\lambda_{lv\sigma} \lambda^*_{ju\sigma} \int_{-\infty}^{+\infty}d\varepsilon \left[ \frac{\Sigma_{\alpha\sigma,nn-n'}^{<}(\varepsilon-n\omega)}{\varepsilon-\Delta_{uj}/\hbar+ i0^{+}}+ 2\pi i\delta (\varepsilon- \Delta_{uj}/\hbar)\Sigma_{\alpha\sigma,nn-n'}^{a}(\varepsilon-n\omega) \right] .
\label{Bar_Rate_Floquet}
\end{split}
\end{equation}
\end{widetext}
where we can identify the relation $\Gamma_{vl,ju}(t)=-\bar{\Gamma}_{lv,uj}^{\ *}(t)$.
The imaginary part of these rates is responsible of a renormalization of the energy spectra of the magnetic impurity, commonly known as Lamb-shift. However, within our wide band approximation these terms exactly cancel in the master equation and current due to the previous relation. By contrast, for the fourth order cotunneling processes, one can demonstrate that the Lamb shift can induce Rabi oscillations even within the wide band approximation \cite{J_Reina_Galvez_2019}.

%% file: main.bbl
\begin{thebibliography}{64}
\expandafter\ifx\csname natexlab\endcsname\relax\def\natexlab#1{#1}\fi
\expandafter\ifx\csname bibnamefont\endcsname\relax
  \def\bibnamefont#1{#1}\fi
\expandafter\ifx\csname bibfnamefont\endcsname\relax
  \def\bibfnamefont#1{#1}\fi
\expandafter\ifx\csname citenamefont\endcsname\relax
  \def\citenamefont#1{#1}\fi
\expandafter\ifx\csname url\endcsname\relax
  \def\url#1{\texttt{#1}}\fi
\expandafter\ifx\csname urlprefix\endcsname\relax\def\urlprefix{URL }\fi
\providecommand{\bibinfo}[2]{#2}
\providecommand{\eprint}[2][]{\url{#2}}

\bibitem[{\citenamefont{der Molen and Liljeroth}(2010)}]{dermolen_charge_2010}
\bibinfo{author}{\bibfnamefont{S.~J.~v.} \bibnamefont{der Molen}}
  \bibnamefont{and}
  \bibinfo{author}{\bibfnamefont{P.}~\bibnamefont{Liljeroth}},
  \bibinfo{journal}{Journal of Physics: Condensed Matter}
  \textbf{\bibinfo{volume}{22}}, \bibinfo{pages}{133001}
  (\bibinfo{year}{2010}), ISSN \bibinfo{issn}{0953-8984},
  \urlprefix\url{http://iopscience.iop.org/0953-8984/22/13/133001}.

\bibitem[{\citenamefont{Evers et~al.}(2020)\citenamefont{Evers, Koryt\'ar,
  Tewari, and van Ruitenbeek}}]{Ferdinand}
\bibinfo{author}{\bibfnamefont{F.}~\bibnamefont{Evers}},
  \bibinfo{author}{\bibfnamefont{R.}~\bibnamefont{Koryt\'ar}},
  \bibinfo{author}{\bibfnamefont{S.}~\bibnamefont{Tewari}}, \bibnamefont{and}
  \bibinfo{author}{\bibfnamefont{J.~M.} \bibnamefont{van Ruitenbeek}},
  \bibinfo{journal}{Rev. Mod. Phys.} \textbf{\bibinfo{volume}{92}},
  \bibinfo{pages}{035001} (\bibinfo{year}{2020}),
  \urlprefix\url{https://link.aps.org/doi/10.1103/RevModPhys.92.035001}.

\bibitem[{\citenamefont{Joachim and Ratner}(2005)}]{joachim_molecular_2005}
\bibinfo{author}{\bibfnamefont{C.}~\bibnamefont{Joachim}} \bibnamefont{and}
  \bibinfo{author}{\bibfnamefont{M.~A.} \bibnamefont{Ratner}},
  \bibinfo{journal}{Proceedings of the National Academy of Sciences}
  \textbf{\bibinfo{volume}{102}}, \bibinfo{pages}{8801} (\bibinfo{year}{2005}),
  ISSN \bibinfo{issn}{0027-8424, 1091-6490},
  \urlprefix\url{http://www.pnas.org/content/102/25/8801}.

\bibitem[{\citenamefont{Abragam and Bleaney}(1970)}]{Abragam_Bleaney_book_1970}
\bibinfo{author}{\bibfnamefont{A.}~\bibnamefont{Abragam}} \bibnamefont{and}
  \bibinfo{author}{\bibfnamefont{B.}~\bibnamefont{Bleaney}},
  \emph{\bibinfo{title}{Electron Paramagnetic Resonance of Transition Ions}}
  (\bibinfo{publisher}{Oxford University Press, Oxford}, \bibinfo{year}{1970}).

\bibitem[{\citenamefont{Fratila and Velders}(2011)}]{Fratila_Velders_arac_2011}
\bibinfo{author}{\bibfnamefont{R.~M.} \bibnamefont{Fratila}} \bibnamefont{and}
  \bibinfo{author}{\bibfnamefont{A.~H.} \bibnamefont{Velders}},
  \bibinfo{journal}{Annual Review of Analytical Chemistry}
  \textbf{\bibinfo{volume}{4}}, \bibinfo{pages}{227} (\bibinfo{year}{2011}),
  \urlprefix\url{https://doi.org/10.1146/annurev-anchem-061010-114024}.

\bibitem[{\citenamefont{Müllegger et~al.}(2014)\citenamefont{Müllegger, Tebi,
  Das, Schöfberger, Faschinger, and Koch}}]{Mullegger}
\bibinfo{author}{\bibfnamefont{S.}~\bibnamefont{Müllegger}},
  \bibinfo{author}{\bibfnamefont{S.}~\bibnamefont{Tebi}},
  \bibinfo{author}{\bibfnamefont{A.~K.} \bibnamefont{Das}},
  \bibinfo{author}{\bibfnamefont{W.}~\bibnamefont{Schöfberger}},
  \bibinfo{author}{\bibfnamefont{F.}~\bibnamefont{Faschinger}},
  \bibnamefont{and} \bibinfo{author}{\bibfnamefont{R.}~\bibnamefont{Koch}},
  \bibinfo{journal}{Physical Review Letters} \textbf{\bibinfo{volume}{113}}
  (\bibinfo{year}{2014}), ISSN \bibinfo{issn}{0031-9007, 1079-7114},
  \urlprefix\url{https://link.aps.org/doi/10.1103/PhysRevLett.113.133001}.

\bibitem[{\citenamefont{Baumann et~al.}(2015)\citenamefont{Baumann, Paul, Choi,
  Lutz, Ardavan, and Heinrich}}]{Baumann_Paul_science_2015}
\bibinfo{author}{\bibfnamefont{S.}~\bibnamefont{Baumann}},
  \bibinfo{author}{\bibfnamefont{W.}~\bibnamefont{Paul}},
  \bibinfo{author}{\bibfnamefont{T.}~\bibnamefont{Choi}},
  \bibinfo{author}{\bibfnamefont{C.~P.} \bibnamefont{Lutz}},
  \bibinfo{author}{\bibfnamefont{A.}~\bibnamefont{Ardavan}}, \bibnamefont{and}
  \bibinfo{author}{\bibfnamefont{A.~J.} \bibnamefont{Heinrich}},
  \bibinfo{journal}{Science} \textbf{\bibinfo{volume}{350}},
  \bibinfo{pages}{417} (\bibinfo{year}{2015}).

\bibitem[{\citenamefont{Delgado and Lorente}(2021)}]{Delgado_Lorente_pss_2021}
\bibinfo{author}{\bibfnamefont{F.}~\bibnamefont{Delgado}} \bibnamefont{and}
  \bibinfo{author}{\bibfnamefont{N.}~\bibnamefont{Lorente}},
  \bibinfo{journal}{Progress in Surface Science} \textbf{\bibinfo{volume}{96}},
  \bibinfo{pages}{100625} (\bibinfo{year}{2021}), ISSN
  \bibinfo{issn}{0079-6816},
  \urlprefix\url{https://www.sciencedirect.com/science/article/pii/S0079681621000137}.

\bibitem[{\citenamefont{Yang et~al.}(2017)\citenamefont{Yang, Bae, Paul,
  Natterer, Willke, Lado, Ferr\'on, Choi, Fern\'andez-Rossier, Heinrich
  et~al.}}]{Yang_Bae_prl_2017}
\bibinfo{author}{\bibfnamefont{K.}~\bibnamefont{Yang}},
  \bibinfo{author}{\bibfnamefont{Y.}~\bibnamefont{Bae}},
  \bibinfo{author}{\bibfnamefont{W.}~\bibnamefont{Paul}},
  \bibinfo{author}{\bibfnamefont{F.~D.} \bibnamefont{Natterer}},
  \bibinfo{author}{\bibfnamefont{P.}~\bibnamefont{Willke}},
  \bibinfo{author}{\bibfnamefont{J.~L.} \bibnamefont{Lado}},
  \bibinfo{author}{\bibfnamefont{A.}~\bibnamefont{Ferr\'on}},
  \bibinfo{author}{\bibfnamefont{T.}~\bibnamefont{Choi}},
  \bibinfo{author}{\bibfnamefont{J.}~\bibnamefont{Fern\'andez-Rossier}},
  \bibinfo{author}{\bibfnamefont{A.~J.} \bibnamefont{Heinrich}},
  \bibnamefont{et~al.}, \bibinfo{journal}{Phys. Rev. Lett.}
  \textbf{\bibinfo{volume}{119}}, \bibinfo{pages}{227206}
  (\bibinfo{year}{2017}),
  \urlprefix\url{https://link.aps.org/doi/10.1103/PhysRevLett.119.227206}.

\bibitem[{\citenamefont{Willke et~al.}(2018{\natexlab{a}})\citenamefont{Willke,
  Paul, Natterer, Yang, Bae, Choi, Fern{\'a}ndez-Rossier, Heinrich, and
  Lutz}}]{Willke_Paul_sciadv_2018}
\bibinfo{author}{\bibfnamefont{P.}~\bibnamefont{Willke}},
  \bibinfo{author}{\bibfnamefont{W.}~\bibnamefont{Paul}},
  \bibinfo{author}{\bibfnamefont{F.~D.} \bibnamefont{Natterer}},
  \bibinfo{author}{\bibfnamefont{K.}~\bibnamefont{Yang}},
  \bibinfo{author}{\bibfnamefont{Y.}~\bibnamefont{Bae}},
  \bibinfo{author}{\bibfnamefont{T.}~\bibnamefont{Choi}},
  \bibinfo{author}{\bibfnamefont{J.}~\bibnamefont{Fern{\'a}ndez-Rossier}},
  \bibinfo{author}{\bibfnamefont{A.~J.} \bibnamefont{Heinrich}},
  \bibnamefont{and} \bibinfo{author}{\bibfnamefont{C.~P.} \bibnamefont{Lutz}},
  \bibinfo{journal}{Science Advances} \textbf{\bibinfo{volume}{4}}
  (\bibinfo{year}{2018}{\natexlab{a}}).

\bibitem[{\citenamefont{Bae et~al.}(2018)\citenamefont{Bae, Yang, Willke, Choi,
  Heinrich, and Lutz}}]{Y_Bae_advanced_science_2018}
\bibinfo{author}{\bibfnamefont{Y.}~\bibnamefont{Bae}},
  \bibinfo{author}{\bibfnamefont{K.}~\bibnamefont{Yang}},
  \bibinfo{author}{\bibfnamefont{P.}~\bibnamefont{Willke}},
  \bibinfo{author}{\bibfnamefont{T.}~\bibnamefont{Choi}},
  \bibinfo{author}{\bibfnamefont{A.~J.} \bibnamefont{Heinrich}},
  \bibnamefont{and} \bibinfo{author}{\bibfnamefont{C.~P.} \bibnamefont{Lutz}},
  \bibinfo{journal}{Science Advances} \textbf{\bibinfo{volume}{4}}
  (\bibinfo{year}{2018}),
  \urlprefix\url{https://advances.sciencemag.org/content/4/11/eaau4159}.

\bibitem[{\citenamefont{Yang et~al.}(2019{\natexlab{a}})\citenamefont{Yang,
  Paul, Phark, Willke, Bae, Choi, Esat, Ardavan, Heinrich, and
  Lutz}}]{yang_coherent_2019}
\bibinfo{author}{\bibfnamefont{K.}~\bibnamefont{Yang}},
  \bibinfo{author}{\bibfnamefont{W.}~\bibnamefont{Paul}},
  \bibinfo{author}{\bibfnamefont{S.-H.} \bibnamefont{Phark}},
  \bibinfo{author}{\bibfnamefont{P.}~\bibnamefont{Willke}},
  \bibinfo{author}{\bibfnamefont{Y.}~\bibnamefont{Bae}},
  \bibinfo{author}{\bibfnamefont{T.}~\bibnamefont{Choi}},
  \bibinfo{author}{\bibfnamefont{T.}~\bibnamefont{Esat}},
  \bibinfo{author}{\bibfnamefont{A.}~\bibnamefont{Ardavan}},
  \bibinfo{author}{\bibfnamefont{A.~J.} \bibnamefont{Heinrich}},
  \bibnamefont{and} \bibinfo{author}{\bibfnamefont{C.~P.} \bibnamefont{Lutz}},
  \bibinfo{journal}{Science} \textbf{\bibinfo{volume}{366}},
  \bibinfo{pages}{509} (\bibinfo{year}{2019}{\natexlab{a}}), ISSN
  \bibinfo{issn}{0036-8075, 1095-9203},
  \urlprefix\url{http://www.sciencemag.org/lookup/doi/10.1126/science.aay6779}.

\bibitem[{\citenamefont{Natterer et~al.}(2017)\citenamefont{Natterer, Yang,
  Paul, Willke, Choi, Greber, Heinrich, and Lutz}}]{Natterer_Yang_nature_2017}
\bibinfo{author}{\bibfnamefont{F.~D.} \bibnamefont{Natterer}},
  \bibinfo{author}{\bibfnamefont{K.}~\bibnamefont{Yang}},
  \bibinfo{author}{\bibfnamefont{W.}~\bibnamefont{Paul}},
  \bibinfo{author}{\bibfnamefont{P.}~\bibnamefont{Willke}},
  \bibinfo{author}{\bibfnamefont{T.}~\bibnamefont{Choi}},
  \bibinfo{author}{\bibfnamefont{T.}~\bibnamefont{Greber}},
  \bibinfo{author}{\bibfnamefont{A.~J.} \bibnamefont{Heinrich}},
  \bibnamefont{and} \bibinfo{author}{\bibfnamefont{C.~P.} \bibnamefont{Lutz}},
  \bibinfo{journal}{Nature} \textbf{\bibinfo{volume}{543}},
  \bibinfo{pages}{226} (\bibinfo{year}{2017}).

\bibitem[{\citenamefont{Choi et~al.}(2017)\citenamefont{Choi, Paul,
  Rolf-Pissarczyk, Macdonald, Natterer, Yang, Willke, Lutz, and
  Heinrich}}]{Choi_Paul_natnano_2017}
\bibinfo{author}{\bibfnamefont{T.}~\bibnamefont{Choi}},
  \bibinfo{author}{\bibfnamefont{W.}~\bibnamefont{Paul}},
  \bibinfo{author}{\bibfnamefont{S.}~\bibnamefont{Rolf-Pissarczyk}},
  \bibinfo{author}{\bibfnamefont{A.~J.} \bibnamefont{Macdonald}},
  \bibinfo{author}{\bibfnamefont{F.~D.} \bibnamefont{Natterer}},
  \bibinfo{author}{\bibfnamefont{K.}~\bibnamefont{Yang}},
  \bibinfo{author}{\bibfnamefont{P.}~\bibnamefont{Willke}},
  \bibinfo{author}{\bibfnamefont{C.~P.} \bibnamefont{Lutz}}, \bibnamefont{and}
  \bibinfo{author}{\bibfnamefont{A.~J.} \bibnamefont{Heinrich}},
  \bibinfo{journal}{Nature nanotechnology} \textbf{\bibinfo{volume}{12}},
  \bibinfo{pages}{420} (\bibinfo{year}{2017}).

\bibitem[{\citenamefont{Willke et~al.}(2018{\natexlab{b}})\citenamefont{Willke,
  Bae, Yang, Lado, Ferr{\'o}n, Choi, Ardavan, Fern{\'a}ndez-Rossier, Heinrich,
  and Lutz}}]{Willke_Bae_science_2018}
\bibinfo{author}{\bibfnamefont{P.}~\bibnamefont{Willke}},
  \bibinfo{author}{\bibfnamefont{Y.}~\bibnamefont{Bae}},
  \bibinfo{author}{\bibfnamefont{K.}~\bibnamefont{Yang}},
  \bibinfo{author}{\bibfnamefont{J.~L.} \bibnamefont{Lado}},
  \bibinfo{author}{\bibfnamefont{A.}~\bibnamefont{Ferr{\'o}n}},
  \bibinfo{author}{\bibfnamefont{T.}~\bibnamefont{Choi}},
  \bibinfo{author}{\bibfnamefont{A.}~\bibnamefont{Ardavan}},
  \bibinfo{author}{\bibfnamefont{J.}~\bibnamefont{Fern{\'a}ndez-Rossier}},
  \bibinfo{author}{\bibfnamefont{A.~J.} \bibnamefont{Heinrich}},
  \bibnamefont{and} \bibinfo{author}{\bibfnamefont{C.~P.} \bibnamefont{Lutz}},
  \bibinfo{journal}{Science} \textbf{\bibinfo{volume}{362}},
  \bibinfo{pages}{336} (\bibinfo{year}{2018}{\natexlab{b}}), ISSN
  \bibinfo{issn}{0036-8075},
  \urlprefix\url{http://science.sciencemag.org/content/362/6412/336}.

\bibitem[{\citenamefont{Willke et~al.}(2019{\natexlab{a}})\citenamefont{Willke,
  Singha, Zhang, Esat, Lutz, Heinrich, and Choi}}]{Willke_Singha_nanolett_2019}
\bibinfo{author}{\bibfnamefont{P.}~\bibnamefont{Willke}},
  \bibinfo{author}{\bibfnamefont{A.}~\bibnamefont{Singha}},
  \bibinfo{author}{\bibfnamefont{X.}~\bibnamefont{Zhang}},
  \bibinfo{author}{\bibfnamefont{T.}~\bibnamefont{Esat}},
  \bibinfo{author}{\bibfnamefont{C.~P.} \bibnamefont{Lutz}},
  \bibinfo{author}{\bibfnamefont{A.~J.} \bibnamefont{Heinrich}},
  \bibnamefont{and} \bibinfo{author}{\bibfnamefont{T.}~\bibnamefont{Choi}},
  \bibinfo{journal}{Nano Letters} \textbf{\bibinfo{volume}{19}},
  \bibinfo{pages}{8201} (\bibinfo{year}{2019}{\natexlab{a}}),
  \bibinfo{note}{pMID: 31661282},
  \urlprefix\url{https://doi.org/10.1021/acs.nanolett.9b03559}.

\bibitem[{\citenamefont{Willke et~al.}(2019{\natexlab{b}})\citenamefont{Willke,
  Yang, Bae, Heinrich, and Lutz}}]{Willke_Yang_natphys_2019}
\bibinfo{author}{\bibfnamefont{P.}~\bibnamefont{Willke}},
  \bibinfo{author}{\bibfnamefont{K.}~\bibnamefont{Yang}},
  \bibinfo{author}{\bibfnamefont{Y.}~\bibnamefont{Bae}},
  \bibinfo{author}{\bibfnamefont{A.}~\bibnamefont{Heinrich}}, \bibnamefont{and}
  \bibinfo{author}{\bibfnamefont{C.}~\bibnamefont{Lutz}},
  \bibinfo{journal}{Nature Physics} \textbf{\bibinfo{volume}{15}},
  \bibinfo{pages}{1005} (\bibinfo{year}{2019}{\natexlab{b}}).

\bibitem[{\citenamefont{Yang et~al.}(2019{\natexlab{b}})\citenamefont{Yang,
  Paul, Natterer, Lado, Bae, Willke, Choi, Ferr\'on, Fern\'andez-Rossier,
  Heinrich et~al.}}]{Yang_Paul_prl_2019}
\bibinfo{author}{\bibfnamefont{K.}~\bibnamefont{Yang}},
  \bibinfo{author}{\bibfnamefont{W.}~\bibnamefont{Paul}},
  \bibinfo{author}{\bibfnamefont{F.~D.} \bibnamefont{Natterer}},
  \bibinfo{author}{\bibfnamefont{J.~L.} \bibnamefont{Lado}},
  \bibinfo{author}{\bibfnamefont{Y.}~\bibnamefont{Bae}},
  \bibinfo{author}{\bibfnamefont{P.}~\bibnamefont{Willke}},
  \bibinfo{author}{\bibfnamefont{T.}~\bibnamefont{Choi}},
  \bibinfo{author}{\bibfnamefont{A.}~\bibnamefont{Ferr\'on}},
  \bibinfo{author}{\bibfnamefont{J.}~\bibnamefont{Fern\'andez-Rossier}},
  \bibinfo{author}{\bibfnamefont{A.~J.} \bibnamefont{Heinrich}},
  \bibnamefont{et~al.}, \bibinfo{journal}{Phys. Rev. Lett.}
  \textbf{\bibinfo{volume}{122}}, \bibinfo{pages}{227203}
  (\bibinfo{year}{2019}{\natexlab{b}}),
  \urlprefix\url{https://link.aps.org/doi/10.1103/PhysRevLett.122.227203}.

\bibitem[{\citenamefont{Seifert et~al.}(2020)\citenamefont{Seifert, Kovarik,
  Juraschek, Spaldin, Gambardella, and Stepanow}}]{Seifert_Kovarik_eabc_2020}
\bibinfo{author}{\bibfnamefont{T.~S.} \bibnamefont{Seifert}},
  \bibinfo{author}{\bibfnamefont{S.}~\bibnamefont{Kovarik}},
  \bibinfo{author}{\bibfnamefont{D.~M.} \bibnamefont{Juraschek}},
  \bibinfo{author}{\bibfnamefont{N.~A.} \bibnamefont{Spaldin}},
  \bibinfo{author}{\bibfnamefont{P.}~\bibnamefont{Gambardella}},
  \bibnamefont{and} \bibinfo{author}{\bibfnamefont{S.}~\bibnamefont{Stepanow}},
  \bibinfo{journal}{Science Advances} \textbf{\bibinfo{volume}{6}}
  (\bibinfo{year}{2020}),
  \urlprefix\url{https://advances.sciencemag.org/content/6/40/eabc5511}.

\bibitem[{\citenamefont{van Weerdenburg et~al.}(2021)\citenamefont{van
  Weerdenburg, Steinbrecher, van Mullekom, Gerritsen, von Allwörden, Natterer,
  and Khajetoorians}}]{Weerdenburg_Steinbrecher_2020}
\bibinfo{author}{\bibfnamefont{W.~M.~J.} \bibnamefont{van Weerdenburg}},
  \bibinfo{author}{\bibfnamefont{M.}~\bibnamefont{Steinbrecher}},
  \bibinfo{author}{\bibfnamefont{N.~P.~E.} \bibnamefont{van Mullekom}},
  \bibinfo{author}{\bibfnamefont{J.~W.} \bibnamefont{Gerritsen}},
  \bibinfo{author}{\bibfnamefont{H.}~\bibnamefont{von Allwörden}},
  \bibinfo{author}{\bibfnamefont{F.~D.} \bibnamefont{Natterer}},
  \bibnamefont{and} \bibinfo{author}{\bibfnamefont{A.~A.}
  \bibnamefont{Khajetoorians}}, \bibinfo{journal}{Review of Scientific
  Instruments} \textbf{\bibinfo{volume}{92}}, \bibinfo{pages}{033906}
  (\bibinfo{year}{2021}), \urlprefix\url{https://doi.org/10.1063/5.0040011}.

\bibitem[{\citenamefont{Steinbrecher et~al.}(2021)\citenamefont{Steinbrecher,
  van Weerdenburg, Walraven, van Mullekom, Gerritsen, Natterer, Badrtdinov,
  Rudenko, Mazurenko, Katsnelson et~al.}}]{Steinbrecher_Weerdenburg_2020}
\bibinfo{author}{\bibfnamefont{M.}~\bibnamefont{Steinbrecher}},
  \bibinfo{author}{\bibfnamefont{W.~M.~J.} \bibnamefont{van Weerdenburg}},
  \bibinfo{author}{\bibfnamefont{E.~F.} \bibnamefont{Walraven}},
  \bibinfo{author}{\bibfnamefont{N.~P.~E.} \bibnamefont{van Mullekom}},
  \bibinfo{author}{\bibfnamefont{J.~W.} \bibnamefont{Gerritsen}},
  \bibinfo{author}{\bibfnamefont{F.~D.} \bibnamefont{Natterer}},
  \bibinfo{author}{\bibfnamefont{D.~I.} \bibnamefont{Badrtdinov}},
  \bibinfo{author}{\bibfnamefont{A.~N.} \bibnamefont{Rudenko}},
  \bibinfo{author}{\bibfnamefont{V.~V.} \bibnamefont{Mazurenko}},
  \bibinfo{author}{\bibfnamefont{M.~I.} \bibnamefont{Katsnelson}},
  \bibnamefont{et~al.}, \bibinfo{journal}{Physical Review B}
  \textbf{\bibinfo{volume}{103}} (\bibinfo{year}{2021}), ISSN
  \bibinfo{issn}{2469-9969},
  \urlprefix\url{http://dx.doi.org/10.1103/PhysRevB.103.155405}.

\bibitem[{\citenamefont{Breuer and
  Petruccione}(2002)}]{Breuer_Petruccione_book_2002}
\bibinfo{author}{\bibfnamefont{H.-P.} \bibnamefont{Breuer}} \bibnamefont{and}
  \bibinfo{author}{\bibfnamefont{F.}~\bibnamefont{Petruccione}},
  \emph{\bibinfo{title}{The theory of open quantum systems}}
  (\bibinfo{publisher}{Oxford University Press}, \bibinfo{year}{2002}).

\bibitem[{\citenamefont{Cohen-Tannoudji
  et~al.}(1998)\citenamefont{Cohen-Tannoudji, Grynberg, and
  Dupont-Roc}}]{Cohen_Grynberg_book_1998}
\bibinfo{author}{\bibfnamefont{C.}~\bibnamefont{Cohen-Tannoudji}},
  \bibinfo{author}{\bibfnamefont{G.}~\bibnamefont{Grynberg}}, \bibnamefont{and}
  \bibinfo{author}{\bibfnamefont{J.}~\bibnamefont{Dupont-Roc}},
  \emph{\bibinfo{title}{Atom-Photon Interactions}} (\bibinfo{publisher}{Wiley
  and Sons, INC., New York}, \bibinfo{year}{1998}).

\bibitem[{\citenamefont{Shakirov et~al.}(2016)\citenamefont{Shakirov,
  Shchadilova, Rubtsov, and Ribeiro}}]{Ribeiro1}
\bibinfo{author}{\bibfnamefont{A.~M.} \bibnamefont{Shakirov}},
  \bibinfo{author}{\bibfnamefont{Y.~E.} \bibnamefont{Shchadilova}},
  \bibinfo{author}{\bibfnamefont{A.~N.} \bibnamefont{Rubtsov}},
  \bibnamefont{and} \bibinfo{author}{\bibfnamefont{P.}~\bibnamefont{Ribeiro}},
  \bibinfo{journal}{Phys. Rev. B} \textbf{\bibinfo{volume}{94}},
  \bibinfo{pages}{224425} (\bibinfo{year}{2016}),
  \urlprefix\url{https://link.aps.org/doi/10.1103/PhysRevB.94.224425}.

\bibitem[{\citenamefont{Nakajima et~al.}(2020)\citenamefont{Nakajima, Noiri,
  Kawasaki, Yoneda, Stano, Amaha, Otsuka, Takeda, Delbecq, Allison
  et~al.}}]{Nakajima_Noiri_prx_2020}
\bibinfo{author}{\bibfnamefont{T.}~\bibnamefont{Nakajima}},
  \bibinfo{author}{\bibfnamefont{A.}~\bibnamefont{Noiri}},
  \bibinfo{author}{\bibfnamefont{K.}~\bibnamefont{Kawasaki}},
  \bibinfo{author}{\bibfnamefont{J.}~\bibnamefont{Yoneda}},
  \bibinfo{author}{\bibfnamefont{P.}~\bibnamefont{Stano}},
  \bibinfo{author}{\bibfnamefont{S.}~\bibnamefont{Amaha}},
  \bibinfo{author}{\bibfnamefont{T.}~\bibnamefont{Otsuka}},
  \bibinfo{author}{\bibfnamefont{K.}~\bibnamefont{Takeda}},
  \bibinfo{author}{\bibfnamefont{M.~R.} \bibnamefont{Delbecq}},
  \bibinfo{author}{\bibfnamefont{G.}~\bibnamefont{Allison}},
  \bibnamefont{et~al.}, \bibinfo{journal}{Phys. Rev. X}
  \textbf{\bibinfo{volume}{10}}, \bibinfo{pages}{011060}
  (\bibinfo{year}{2020}),
  \urlprefix\url{https://link.aps.org/doi/10.1103/PhysRevX.10.011060}.

\bibitem[{\citenamefont{Yang et~al.}(2019{\natexlab{c}})\citenamefont{Yang,
  Coppersmith, and Friesen}}]{yang2019achieving}
\bibinfo{author}{\bibfnamefont{Y.-C.} \bibnamefont{Yang}},
  \bibinfo{author}{\bibfnamefont{S.}~\bibnamefont{Coppersmith}},
  \bibnamefont{and} \bibinfo{author}{\bibfnamefont{M.}~\bibnamefont{Friesen}},
  \bibinfo{journal}{npj Quantum Information} \textbf{\bibinfo{volume}{5}},
  \bibinfo{pages}{1} (\bibinfo{year}{2019}{\natexlab{c}}).

\bibitem[{\citenamefont{Vandersypen et~al.}(2017)\citenamefont{Vandersypen,
  Bluhm, Clarke, Dzurak, Ishihara, Morello, Reilly, Schreiber, and
  Veldhorst}}]{vandersypen2017interfacing}
\bibinfo{author}{\bibfnamefont{L.}~\bibnamefont{Vandersypen}},
  \bibinfo{author}{\bibfnamefont{H.}~\bibnamefont{Bluhm}},
  \bibinfo{author}{\bibfnamefont{J.}~\bibnamefont{Clarke}},
  \bibinfo{author}{\bibfnamefont{A.}~\bibnamefont{Dzurak}},
  \bibinfo{author}{\bibfnamefont{R.}~\bibnamefont{Ishihara}},
  \bibinfo{author}{\bibfnamefont{A.}~\bibnamefont{Morello}},
  \bibinfo{author}{\bibfnamefont{D.}~\bibnamefont{Reilly}},
  \bibinfo{author}{\bibfnamefont{L.}~\bibnamefont{Schreiber}},
  \bibnamefont{and}
  \bibinfo{author}{\bibfnamefont{M.}~\bibnamefont{Veldhorst}},
  \bibinfo{journal}{npj Quantum Information} \textbf{\bibinfo{volume}{3}},
  \bibinfo{pages}{1} (\bibinfo{year}{2017}).

\bibitem[{\citenamefont{Morello et~al.}(2020)\citenamefont{Morello, Pla,
  Bertet, and Jamieson}}]{Morello_Pla_aqt_2020}
\bibinfo{author}{\bibfnamefont{A.}~\bibnamefont{Morello}},
  \bibinfo{author}{\bibfnamefont{J.~J.} \bibnamefont{Pla}},
  \bibinfo{author}{\bibfnamefont{P.}~\bibnamefont{Bertet}}, \bibnamefont{and}
  \bibinfo{author}{\bibfnamefont{D.~N.} \bibnamefont{Jamieson}},
  \bibinfo{journal}{Advanced Quantum Technologies}
  \textbf{\bibinfo{volume}{3}}, \bibinfo{pages}{2000005}
  (\bibinfo{year}{2020}),
  \urlprefix\url{https://onlinelibrary.wiley.com/doi/abs/10.1002/qute.202000005}.

\bibitem[{\citenamefont{Engel and Loss}(2001)}]{Engel_Hans_prl_2001}
\bibinfo{author}{\bibfnamefont{H.-A.} \bibnamefont{Engel}} \bibnamefont{and}
  \bibinfo{author}{\bibfnamefont{D.}~\bibnamefont{Loss}},
  \bibinfo{journal}{Phys. Rev. Lett.} \textbf{\bibinfo{volume}{86}},
  \bibinfo{pages}{4648} (\bibinfo{year}{2001}),
  \urlprefix\url{https://link.aps.org/doi/10.1103/PhysRevLett.86.4648}.

\bibitem[{\citenamefont{Engel and Loss}(2002)}]{Engel_Loss_prb_2002}
\bibinfo{author}{\bibfnamefont{H.-A.} \bibnamefont{Engel}} \bibnamefont{and}
  \bibinfo{author}{\bibfnamefont{D.}~\bibnamefont{Loss}},
  \bibinfo{journal}{Phys. Rev. B} \textbf{\bibinfo{volume}{65}},
  \bibinfo{pages}{195321} (\bibinfo{year}{2002}),
  \urlprefix\url{https://link.aps.org/doi/10.1103/PhysRevB.65.195321}.

\bibitem[{\citenamefont{Shakirov et~al.}(2019)\citenamefont{Shakirov, Rubtsov,
  and Ribeiro}}]{Ribeiro2}
\bibinfo{author}{\bibfnamefont{A.~M.} \bibnamefont{Shakirov}},
  \bibinfo{author}{\bibfnamefont{A.~N.} \bibnamefont{Rubtsov}},
  \bibnamefont{and} \bibinfo{author}{\bibfnamefont{P.}~\bibnamefont{Ribeiro}},
  \bibinfo{journal}{Phys. Rev. B} \textbf{\bibinfo{volume}{99}},
  \bibinfo{pages}{054434} (\bibinfo{year}{2019}),
  \urlprefix\url{https://link.aps.org/doi/10.1103/PhysRevB.99.054434}.

\bibitem[{\citenamefont{Cavaliere
  et~al.}(2009{\natexlab{a}})\citenamefont{Cavaliere, Governale, and
  K\"onig}}]{Cavaliere_2009}
\bibinfo{author}{\bibfnamefont{F.}~\bibnamefont{Cavaliere}},
  \bibinfo{author}{\bibfnamefont{M.}~\bibnamefont{Governale}},
  \bibnamefont{and} \bibinfo{author}{\bibfnamefont{J.}~\bibnamefont{K\"onig}},
  \bibinfo{journal}{Phys. Rev. Lett.} \textbf{\bibinfo{volume}{103}},
  \bibinfo{pages}{136801} (\bibinfo{year}{2009}{\natexlab{a}}),
  \urlprefix\url{https://link.aps.org/doi/10.1103/PhysRevLett.103.136801}.

\bibitem[{\citenamefont{Cavaliere
  et~al.}(2009{\natexlab{b}})\citenamefont{Cavaliere, Governale, and
  K\"onig}}]{cava-09}
\bibinfo{author}{\bibfnamefont{F.}~\bibnamefont{Cavaliere}},
  \bibinfo{author}{\bibfnamefont{M.}~\bibnamefont{Governale}},
  \bibnamefont{and} \bibinfo{author}{\bibfnamefont{J.}~\bibnamefont{K\"onig}},
  \bibinfo{journal}{Phys. Rev. Lett.} \textbf{\bibinfo{volume}{103}},
  \bibinfo{pages}{136801} (\bibinfo{year}{2009}{\natexlab{b}}),
  \urlprefix\url{https://link.aps.org/doi/10.1103/PhysRevLett.103.136801}.

\bibitem[{\citenamefont{Delgado and
  Fernández-Rossier}(2017)}]{Delgado_Rossier_pss_2017}
\bibinfo{author}{\bibfnamefont{F.}~\bibnamefont{Delgado}} \bibnamefont{and}
  \bibinfo{author}{\bibfnamefont{J.}~\bibnamefont{Fernández-Rossier}},
  \bibinfo{journal}{Progress in Surface Science} \textbf{\bibinfo{volume}{92}},
  \bibinfo{pages}{40} (\bibinfo{year}{2017}), ISSN \bibinfo{issn}{0079-6816},
  \urlprefix\url{https://www.sciencedirect.com/science/article/pii/S0079681616300351}.

\bibitem[{\citenamefont{Grifoni and Hänggi}(1998)}]{Grifoni_Hanggi_2021}
\bibinfo{author}{\bibfnamefont{M.}~\bibnamefont{Grifoni}} \bibnamefont{and}
  \bibinfo{author}{\bibfnamefont{P.}~\bibnamefont{Hänggi}},
  \bibinfo{journal}{Physics Reports} \textbf{\bibinfo{volume}{304}},
  \bibinfo{pages}{229} (\bibinfo{year}{1998}), ISSN \bibinfo{issn}{03701573},
  \urlprefix\url{https://linkinghub.elsevier.com/retrieve/pii/S0370157398000222}.

\bibitem[{\citenamefont{Rammer}(2007)}]{rammer_2007}
\bibinfo{author}{\bibfnamefont{J.}~\bibnamefont{Rammer}},
  \emph{\bibinfo{title}{Quantum Field Theory of Non-equilibrium States}}
  (\bibinfo{publisher}{Cambridge University Press}, \bibinfo{year}{2007}).

\bibitem[{\citenamefont{Jauho}(2006)}]{A_P_Jauho_2006}
\bibinfo{author}{\bibfnamefont{A.~P.} \bibnamefont{Jauho}},
  \emph{\bibinfo{title}{Introduction to the keldysh nonequilibrium green
  function technique}} (\bibinfo{year}{2006}),
  \urlprefix\url{https://nanohub.org/resources/1877}.

\bibitem[{\citenamefont{Pastawski}(1992)}]{pastawski1992classical}
\bibinfo{author}{\bibfnamefont{H.~M.} \bibnamefont{Pastawski}},
  \bibinfo{journal}{Phys. Rev. B} \textbf{\bibinfo{volume}{46}},
  \bibinfo{pages}{4053} (\bibinfo{year}{1992}),
  \urlprefix\url{https://link.aps.org/doi/10.1103/PhysRevB.46.4053}.

\bibitem[{\citenamefont{Jauho et~al.}(1994)\citenamefont{Jauho, Wingreen, and
  Meir}}]{Jauho_Wingreen_1994}
\bibinfo{author}{\bibfnamefont{A.-P.} \bibnamefont{Jauho}},
  \bibinfo{author}{\bibfnamefont{N.~S.} \bibnamefont{Wingreen}},
  \bibnamefont{and} \bibinfo{author}{\bibfnamefont{Y.}~\bibnamefont{Meir}},
  \bibinfo{journal}{Phys. Rev. B} \textbf{\bibinfo{volume}{50}},
  \bibinfo{pages}{5528} (\bibinfo{year}{1994}),
  \urlprefix\url{https://link.aps.org/doi/10.1103/PhysRevB.50.5528}.

\bibitem[{\citenamefont{Wingreen et~al.}(1993)\citenamefont{Wingreen, Jauho,
  and Meir}}]{Wingreen}
\bibinfo{author}{\bibfnamefont{N.~S.} \bibnamefont{Wingreen}},
  \bibinfo{author}{\bibfnamefont{A.-P.} \bibnamefont{Jauho}}, \bibnamefont{and}
  \bibinfo{author}{\bibfnamefont{Y.}~\bibnamefont{Meir}},
  \bibinfo{journal}{Phys. Rev. B} \textbf{\bibinfo{volume}{48}},
  \bibinfo{pages}{8487} (\bibinfo{year}{1993}),
  \urlprefix\url{https://link.aps.org/doi/10.1103/PhysRevB.48.8487}.

\bibitem[{\citenamefont{Arrachea}(2005)}]{arrachea_green-function_2005}
\bibinfo{author}{\bibfnamefont{L.}~\bibnamefont{Arrachea}},
  \bibinfo{journal}{Physical Review B} \textbf{\bibinfo{volume}{72}},
  \bibinfo{pages}{125349} (\bibinfo{year}{2005}),
  \urlprefix\url{https://link.aps.org/doi/10.1103/PhysRevB.72.125349}.

\bibitem[{\citenamefont{Moskalets and Büttiker}(2002)}]{Moskalets_2002}
\bibinfo{author}{\bibfnamefont{M.}~\bibnamefont{Moskalets}} \bibnamefont{and}
  \bibinfo{author}{\bibfnamefont{M.}~\bibnamefont{Büttiker}},
  \bibinfo{journal}{Physical Review B} \textbf{\bibinfo{volume}{66}}
  (\bibinfo{year}{2002}), ISSN \bibinfo{issn}{1095-3795},
  \urlprefix\url{http://dx.doi.org/10.1103/PhysRevB.66.205320}.

\bibitem[{\citenamefont{Arrachea and
  Moskalets}(2006)}]{arrachea-moskalets-2006}
\bibinfo{author}{\bibfnamefont{L.}~\bibnamefont{Arrachea}} \bibnamefont{and}
  \bibinfo{author}{\bibfnamefont{M.}~\bibnamefont{Moskalets}},
  \bibinfo{journal}{Phys. Rev. B} \textbf{\bibinfo{volume}{74}},
  \bibinfo{pages}{245322} (\bibinfo{year}{2006}),
  \urlprefix\url{https://link.aps.org/doi/10.1103/PhysRevB.74.245322}.

\bibitem[{\citenamefont{Schoeller and Schön}(1994)}]{Schoeller}
\bibinfo{author}{\bibfnamefont{H.}~\bibnamefont{Schoeller}} \bibnamefont{and}
  \bibinfo{author}{\bibfnamefont{G.}~\bibnamefont{Schön}},
  \bibinfo{journal}{Physical Review B} \textbf{\bibinfo{volume}{50}},
  \bibinfo{pages}{18436} (\bibinfo{year}{1994}), \bibinfo{note}{publisher:
  American Physical Society},
  \urlprefix\url{https://link.aps.org/doi/10.1103/PhysRevB.50.18436}.

\bibitem[{\citenamefont{Schoeller and Sch\"on}(1994)}]{schoen-94}
\bibinfo{author}{\bibfnamefont{H.}~\bibnamefont{Schoeller}} \bibnamefont{and}
  \bibinfo{author}{\bibfnamefont{G.}~\bibnamefont{Sch\"on}},
  \bibinfo{journal}{Phys. Rev. B} \textbf{\bibinfo{volume}{50}},
  \bibinfo{pages}{18436} (\bibinfo{year}{1994}),
  \urlprefix\url{https://link.aps.org/doi/10.1103/PhysRevB.50.18436}.

\bibitem[{\citenamefont{K\"onig
  et~al.}(1996{\natexlab{a}})\citenamefont{K\"onig, Schoeller, and
  Sch\"on}}]{koenig-96-1}
\bibinfo{author}{\bibfnamefont{J.}~\bibnamefont{K\"onig}},
  \bibinfo{author}{\bibfnamefont{H.}~\bibnamefont{Schoeller}},
  \bibnamefont{and} \bibinfo{author}{\bibfnamefont{G.}~\bibnamefont{Sch\"on}},
  \bibinfo{journal}{Phys. Rev. Lett.} \textbf{\bibinfo{volume}{76}},
  \bibinfo{pages}{1715} (\bibinfo{year}{1996}{\natexlab{a}}),
  \urlprefix\url{https://link.aps.org/doi/10.1103/PhysRevLett.76.1715}.

\bibitem[{\citenamefont{K\"onig
  et~al.}(1996{\natexlab{b}})\citenamefont{K\"onig, Schmid, Schoeller, and
  Sch\"on}}]{koenig-96-2}
\bibinfo{author}{\bibfnamefont{J.}~\bibnamefont{K\"onig}},
  \bibinfo{author}{\bibfnamefont{J.}~\bibnamefont{Schmid}},
  \bibinfo{author}{\bibfnamefont{H.}~\bibnamefont{Schoeller}},
  \bibnamefont{and} \bibinfo{author}{\bibfnamefont{G.}~\bibnamefont{Sch\"on}},
  \bibinfo{journal}{Phys. Rev. B} \textbf{\bibinfo{volume}{54}},
  \bibinfo{pages}{16820} (\bibinfo{year}{1996}{\natexlab{b}}),
  \urlprefix\url{https://link.aps.org/doi/10.1103/PhysRevB.54.16820}.

\bibitem[{\citenamefont{Esposito and
  Galperin}(2009)}]{Esposito_Galperin_prb_2009}
\bibinfo{author}{\bibfnamefont{M.}~\bibnamefont{Esposito}} \bibnamefont{and}
  \bibinfo{author}{\bibfnamefont{M.}~\bibnamefont{Galperin}},
  \bibinfo{journal}{Phys. Rev. B} \textbf{\bibinfo{volume}{79}},
  \bibinfo{pages}{205303} (\bibinfo{year}{2009}).

\bibitem[{\citenamefont{Splettstoesser
  et~al.}(2006)\citenamefont{Splettstoesser, Governale, K\"onig, and
  Fazio}}]{spletts-06}
\bibinfo{author}{\bibfnamefont{J.}~\bibnamefont{Splettstoesser}},
  \bibinfo{author}{\bibfnamefont{M.}~\bibnamefont{Governale}},
  \bibinfo{author}{\bibfnamefont{J.}~\bibnamefont{K\"onig}}, \bibnamefont{and}
  \bibinfo{author}{\bibfnamefont{R.}~\bibnamefont{Fazio}},
  \bibinfo{journal}{Phys. Rev. B} \textbf{\bibinfo{volume}{74}},
  \bibinfo{pages}{085305} (\bibinfo{year}{2006}),
  \urlprefix\url{https://link.aps.org/doi/10.1103/PhysRevB.74.085305}.

\bibitem[{\citenamefont{Bhandari et~al.}(2021)\citenamefont{Bhandari, Fazio,
  Taddei, and Arrachea}}]{Bibek_Bhandari_2021_nonequilibrium}
\bibinfo{author}{\bibfnamefont{B.}~\bibnamefont{Bhandari}},
  \bibinfo{author}{\bibfnamefont{R.}~\bibnamefont{Fazio}},
  \bibinfo{author}{\bibfnamefont{F.}~\bibnamefont{Taddei}}, \bibnamefont{and}
  \bibinfo{author}{\bibfnamefont{L.}~\bibnamefont{Arrachea}},
  \bibinfo{journal}{Phys. Rev. B} \textbf{\bibinfo{volume}{104}},
  \bibinfo{pages}{035425} (\bibinfo{year}{2021}),
  \urlprefix\url{https://link.aps.org/doi/10.1103/PhysRevB.104.035425}.

\bibitem[{\citenamefont{Veldman et~al.}(2021)\citenamefont{Veldman, Farinacci,
  Rejali, Broekhoven, Gobeil, Coffey, Ternes, and Otte}}]{Veldman}
\bibinfo{author}{\bibfnamefont{L.~M.} \bibnamefont{Veldman}},
  \bibinfo{author}{\bibfnamefont{L.}~\bibnamefont{Farinacci}},
  \bibinfo{author}{\bibfnamefont{R.}~\bibnamefont{Rejali}},
  \bibinfo{author}{\bibfnamefont{R.}~\bibnamefont{Broekhoven}},
  \bibinfo{author}{\bibfnamefont{J.}~\bibnamefont{Gobeil}},
  \bibinfo{author}{\bibfnamefont{D.}~\bibnamefont{Coffey}},
  \bibinfo{author}{\bibfnamefont{M.}~\bibnamefont{Ternes}}, \bibnamefont{and}
  \bibinfo{author}{\bibfnamefont{A.~F.} \bibnamefont{Otte}},
  \bibinfo{journal}{Science} \textbf{\bibinfo{volume}{372}},
  \bibinfo{pages}{964} (\bibinfo{year}{2021}), \bibinfo{note}{publisher:
  American Association for the Advancement of Science},
  \urlprefix\url{https://www.science.org/doi/full/10.1126/science.abg8223}.

\bibitem[{\citenamefont{Phark et~al.}(2021)\citenamefont{Phark, Chen, Wolf,
  Bui, Wang, Haze, Kim, Lutz, Heinrich, and Bae}}]{Phark}
\bibinfo{author}{\bibfnamefont{S.-H.} \bibnamefont{Phark}},
  \bibinfo{author}{\bibfnamefont{Y.}~\bibnamefont{Chen}},
  \bibinfo{author}{\bibfnamefont{C.}~\bibnamefont{Wolf}},
  \bibinfo{author}{\bibfnamefont{H.~T.} \bibnamefont{Bui}},
  \bibinfo{author}{\bibfnamefont{Y.}~\bibnamefont{Wang}},
  \bibinfo{author}{\bibfnamefont{M.}~\bibnamefont{Haze}},
  \bibinfo{author}{\bibfnamefont{J.}~\bibnamefont{Kim}},
  \bibinfo{author}{\bibfnamefont{C.~P.} \bibnamefont{Lutz}},
  \bibinfo{author}{\bibfnamefont{A.~J.} \bibnamefont{Heinrich}},
  \bibnamefont{and} \bibinfo{author}{\bibfnamefont{Y.}~\bibnamefont{Bae}},
  \bibinfo{journal}{arXiv:2108.09880 [cond-mat, physics:quant-ph]}
  (\bibinfo{year}{2021}), \urlprefix\url{http://arxiv.org/abs/2108.09880}.

\bibitem[{Note1()}]{Note1}
Note1, \bibinfo{note}{by contrast, in STM-ESR, a cotunneling regime applies
  where the impurity's orbital lies out of the bias window and the current
  proceeds through the resonance's tails caused by the couplings to the
  electrodes \cite {Delgado_Rossier_prb_2011,J_Reina_Galvez_2019}. It would
  require to go to higher order in the hopping.}

\bibitem[{\citenamefont{Choi et~al.}(2019)\citenamefont{Choi, Lorente, Wiebe,
  von Bergmann, Otte, and Heinrich}}]{spinchains}
\bibinfo{author}{\bibfnamefont{D.-J.} \bibnamefont{Choi}},
  \bibinfo{author}{\bibfnamefont{N.}~\bibnamefont{Lorente}},
  \bibinfo{author}{\bibfnamefont{J.}~\bibnamefont{Wiebe}},
  \bibinfo{author}{\bibfnamefont{K.}~\bibnamefont{von Bergmann}},
  \bibinfo{author}{\bibfnamefont{A.~F.} \bibnamefont{Otte}}, \bibnamefont{and}
  \bibinfo{author}{\bibfnamefont{A.~J.} \bibnamefont{Heinrich}},
  \bibinfo{journal}{Reviews of Modern Physics} \textbf{\bibinfo{volume}{91}},
  \bibinfo{pages}{041001} (\bibinfo{year}{2019}), \bibinfo{note}{publisher:
  American Physical Society},
  \urlprefix\url{https://link.aps.org/doi/10.1103/RevModPhys.91.041001}.

\bibitem[{\citenamefont{Reina~G\'alvez
  et~al.}(2019)\citenamefont{Reina~G\'alvez, Wolf, Delgado, and
  Lorente}}]{J_Reina_Galvez_2019}
\bibinfo{author}{\bibfnamefont{J.}~\bibnamefont{Reina~G\'alvez}},
  \bibinfo{author}{\bibfnamefont{C.}~\bibnamefont{Wolf}},
  \bibinfo{author}{\bibfnamefont{F.}~\bibnamefont{Delgado}}, \bibnamefont{and}
  \bibinfo{author}{\bibfnamefont{N.}~\bibnamefont{Lorente}},
  \bibinfo{journal}{Phys. Rev. B} \textbf{\bibinfo{volume}{100}},
  \bibinfo{pages}{035411} (\bibinfo{year}{2019}),
  \urlprefix\url{https://link.aps.org/doi/10.1103/PhysRevB.100.035411}.

\bibitem[{\citenamefont{Wolf et~al.}(2020)\citenamefont{Wolf, Delgado, Reina,
  and Lorente}}]{Wolf_C_and_Delgado_F_2020}
\bibinfo{author}{\bibfnamefont{C.}~\bibnamefont{Wolf}},
  \bibinfo{author}{\bibfnamefont{F.}~\bibnamefont{Delgado}},
  \bibinfo{author}{\bibfnamefont{J.}~\bibnamefont{Reina}}, \bibnamefont{and}
  \bibinfo{author}{\bibfnamefont{N.}~\bibnamefont{Lorente}},
  \bibinfo{journal}{The Journal of Physical Chemistry A}
  \textbf{\bibinfo{volume}{124}}, \bibinfo{pages}{2318} (\bibinfo{year}{2020}),
  \bibinfo{note}{pMID: 32098473},
  \urlprefix\url{https://doi.org/10.1021/acs.jpca.9b10749}.

\bibitem[{\citenamefont{Dorn et~al.}(2021)\citenamefont{Dorn, Arrigoni, and
  von~der Linden}}]{Dorn_2021}
\bibinfo{author}{\bibfnamefont{G.}~\bibnamefont{Dorn}},
  \bibinfo{author}{\bibfnamefont{E.}~\bibnamefont{Arrigoni}}, \bibnamefont{and}
  \bibinfo{author}{\bibfnamefont{W.}~\bibnamefont{von~der Linden}},
  \bibinfo{journal}{Journal of Physics A: Mathematical and Theoretical}
  \textbf{\bibinfo{volume}{54}}, \bibinfo{pages}{075301}
  (\bibinfo{year}{2021}),
  \urlprefix\url{https://doi.org/10.1088/1751-8121/abd736}.

\bibitem[{\citenamefont{Kim et~al.}(2021)\citenamefont{Kim, Jang, Bui, Choi,
  Wolf, Delgado, Krylov, Lee, Yoon, Lutz et~al.}}]{kim2021spin}
\bibinfo{author}{\bibfnamefont{J.}~\bibnamefont{Kim}},
  \bibinfo{author}{\bibfnamefont{W.-j.} \bibnamefont{Jang}},
  \bibinfo{author}{\bibfnamefont{T.~H.} \bibnamefont{Bui}},
  \bibinfo{author}{\bibfnamefont{D.-j.} \bibnamefont{Choi}},
  \bibinfo{author}{\bibfnamefont{C.}~\bibnamefont{Wolf}},
  \bibinfo{author}{\bibfnamefont{F.}~\bibnamefont{Delgado}},
  \bibinfo{author}{\bibfnamefont{D.}~\bibnamefont{Krylov}},
  \bibinfo{author}{\bibfnamefont{S.}~\bibnamefont{Lee}},
  \bibinfo{author}{\bibfnamefont{S.}~\bibnamefont{Yoon}},
  \bibinfo{author}{\bibfnamefont{C.~P.} \bibnamefont{Lutz}},
  \bibnamefont{et~al.}, \bibinfo{journal}{arXiv preprint arXiv:2103.09582}
  (\bibinfo{year}{2021}).

\bibitem[{Note2()}]{Note2}
Note2, \bibinfo{note}{the limit $B_x \rightarrow 0$ has to be taken with care,
  but confirms the above explanation.}

\bibitem[{\citenamefont{Caso et~al.}(2014)\citenamefont{Caso, Horovitz, and
  Arrachea}}]{caso_model_2014}
\bibinfo{author}{\bibfnamefont{A.}~\bibnamefont{Caso}},
  \bibinfo{author}{\bibfnamefont{B.}~\bibnamefont{Horovitz}}, \bibnamefont{and}
  \bibinfo{author}{\bibfnamefont{L.}~\bibnamefont{Arrachea}},
  \bibinfo{journal}{Physical Review B} \textbf{\bibinfo{volume}{89}}
  (\bibinfo{year}{2014}), ISSN \bibinfo{issn}{1098-0121, 1550-235X},
  \urlprefix\url{https://link.aps.org/doi/10.1103/PhysRevB.89.075412}.

\bibitem[{\citenamefont{Entin-Wohlman
  et~al.}(2020{\natexlab{a}})\citenamefont{Entin-Wohlman, Shekhter, Jonson, and
  Aharony}}]{Entin-Wohlman_2020_spin-orbit}
\bibinfo{author}{\bibfnamefont{O.}~\bibnamefont{Entin-Wohlman}},
  \bibinfo{author}{\bibfnamefont{R.~I.} \bibnamefont{Shekhter}},
  \bibinfo{author}{\bibfnamefont{M.}~\bibnamefont{Jonson}}, \bibnamefont{and}
  \bibinfo{author}{\bibfnamefont{A.}~\bibnamefont{Aharony}},
  \bibinfo{journal}{Phys. Rev. B} \textbf{\bibinfo{volume}{101}},
  \bibinfo{pages}{121303} (\bibinfo{year}{2020}{\natexlab{a}}),
  \urlprefix\url{https://link.aps.org/doi/10.1103/PhysRevB.101.121303}.

\bibitem[{\citenamefont{Entin-Wohlman
  et~al.}(2020{\natexlab{b}})\citenamefont{Entin-Wohlman, Shekhter, Jonson, and
  Aharony}}]{Entin-Wohlman_2020_Rashba}
\bibinfo{author}{\bibfnamefont{O.}~\bibnamefont{Entin-Wohlman}},
  \bibinfo{author}{\bibfnamefont{R.~I.} \bibnamefont{Shekhter}},
  \bibinfo{author}{\bibfnamefont{M.}~\bibnamefont{Jonson}}, \bibnamefont{and}
  \bibinfo{author}{\bibfnamefont{A.}~\bibnamefont{Aharony}},
  \bibinfo{journal}{Phys. Rev. B} \textbf{\bibinfo{volume}{102}},
  \bibinfo{pages}{075419} (\bibinfo{year}{2020}{\natexlab{b}}),
  \urlprefix\url{https://link.aps.org/doi/10.1103/PhysRevB.102.075419}.

\bibitem[{\citenamefont{Hewson}(1997)}]{Hewson_book_1997}
\bibinfo{author}{\bibfnamefont{A.~C.} \bibnamefont{Hewson}},
  \emph{\bibinfo{title}{The Kondo Problem to Heavy Fermions}}
  (\bibinfo{publisher}{Cambridge University Press},
  \bibinfo{address}{Cambridge, UK}, \bibinfo{year}{1997}).

\bibitem[{\citenamefont{Delgado and
  Fern\'andez-Rossier}(2011)}]{Delgado_Rossier_prb_2011}
\bibinfo{author}{\bibfnamefont{F.}~\bibnamefont{Delgado}} \bibnamefont{and}
  \bibinfo{author}{\bibfnamefont{J.}~\bibnamefont{Fern\'andez-Rossier}},
  \bibinfo{journal}{Phys. Rev. B} \textbf{\bibinfo{volume}{84}},
  \bibinfo{pages}{045439} (\bibinfo{year}{2011}),
  \urlprefix\url{https://link.aps.org/doi/10.1103/PhysRevB.84.045439}.

\end{thebibliography}
